\documentclass[review,3p,times]{elsarticle}

\usepackage[utf8]{inputenc}
\usepackage[T1]{fontenc}
\usepackage{graphicx}
\usepackage{subcaption}
\usepackage{booktabs}
\usepackage{hyperref}
\usepackage{url}
\usepackage{amsfonts}
\usepackage{nicefrac}
\usepackage{microtype}
\usepackage{xcolor}
\usepackage{amsmath}
\usepackage{amsthm}
\usepackage{algorithm}
\usepackage{algpseudocode}
\usepackage{threeparttable}

\newcommand\norm[1]{\left\lVert#1\right\rVert}
\newcommand\aaa{\alpha}
\newcommand\bbb{\beta}     

\biboptions{sort&compress}

\journal{Computer Methods and Programs in Biomedicine}

\begin{document}

\begin{frontmatter}

%% ── title ───────────────────────────────────────────────────────────
\title{SDFStent: Real-time interactive virtual stenting via SDF deformation fields}

%% ── authors & affiliations (elsarticle syntax) ─────────────────────
\author[icme]{Bohan J. Li}
\author[cheme]{Nicholas C. Dorn}
\author[queens]{Andras Lasso}
\author[chopaccm,chop]{Matthew A. Jolley}
\author[ped,bioe]{Jeffrey A. Feinstein}
\author[icme,cs]{Doug L. James}
\author[icme,ped,bioe]{Alison~L.~Marsden\corref{cor1}}
\ead{amarsden@stanford.edu}
\cortext[cor1]{Corresponding author. Tel.: +1 650 723 7739.}

\affiliation[icme]{organization={Institute for Computational and
            Mathematical Engineering, Stanford University},
            city={Stanford},
            state={CA},
            postcode={94305},
            country={United States}}
\affiliation[cheme]{organization={Department of Chemical Engineering,
            Stanford University},
            city={Stanford},
            state={CA},
            postcode={94305},
            country={United States}}
\affiliation[queens]{organization={Laboratory for Percutaneous Surgery,
            Queen's University},
            city={Kingston},
            state={ON},
            postcode={K7L 2N8},
            country={Canada}}
\affiliation[chopaccm]{organization={Department of Anesthesiology and Critical Care Medicine,    
            Children’s Hospital of Philadelphia},
            city={Philadelphia},
            state={PA},
            postcode={19104},
            country={United States}}
\affiliation[chop]{organization={Division of Cardiology, Department of Pediatrics,    
            Children’s Hospital of Philadelphia},
            city={Philadelphia},
            state={PA},
            postcode={19104},
            country={United States}}
\affiliation[ped]{organization={Division of Cardiology, Department of Pediatrics,
            Stanford University School of Medicine},
            city={Palo Alto},
            state={CA},
            postcode={94305},
            country={United States}}
\affiliation[cs]{organization={Department of Computer Science,
            Stanford University},
            city={Stanford},
            state={CA},
            postcode={94305},
            country={United States}}
\affiliation[bioe]{organization={Department of Bioengineering,
            Stanford University},
            city={Stanford},
            state={CA},
            postcode={94305},
            country={United States}}

%% ── structured abstract (CMPB house style) ─────────────────────────
\begin{abstract}

\noindent\textbf{Background and Objective:} Stenting is among the most common transcatheter interventions for congenital heart disease (CHD). Patient-specific computational fluid dynamics (CFD) simulations can predict hemodynamic outcomes of intervention scenarios but require post-operative vascular geometries that reflect stent-induced shape changes, which existing tools either model inadequately or require extensive time or manual effort to generate. We present SDFStent, a signed distance function (SDF) based mesh deformation method for virtual stenting that operates in real time, maintains mesh integrity, and preserves junction geometry.

\noindent\textbf{Methods:} The stent is modeled as a pipe surface composed of piecewise-capsule SDFs joined by a smooth-minimum operator. Mesh vertices near the expanding SDF surface are displaced along the SDF gradient with a compactly supported fall-off function and an alpha blending mask. SDFStent was benchmarked against three existing approaches and validated on three tetralogy of Fallot (ToF) patients and three coarctation of the aorta (CoA) patients using rigid-wall steady-state CFD simulations against clinical catheterization measurements.

\noindent\textbf{Results:} Against a prescribed diameter of 6.0 mm, the method produced a mean stented diameter of 5.92 $\pm$ 0.08 mm in 1.5 s, over 100$\times$ faster than the best stenting-specific comparator. All output meshes were watertight and self-intersection-free. CFD-simulated post-operative pressure drops agreed with clinical measurements within 4 mmHg (mean error 2 mmHg).

\noindent\textbf{Conclusions:} SDFStent produces simulation-ready post-stent models that match prescribed stent dimensions at interactive speeds, from pre-operative anatomy and catheterization data alone. The implementation is open-source and available in 3D Slicer. Its scriptable architecture enables automated generation of large synthetic cohorts for data-driven surrogate modeling.

\end{abstract}

%% ── keywords ────────────────────────────────────────────────────────

\end{frontmatter}

%% =====================================================================
%%  1.  INTRODUCTION
%% =====================================================================
\section{Introduction}
\label{sec:introduction}

Congenital heart defects (CHDs) affect roughly 1\% of newborns and remain a leading cause of infant mortality and childhood morbidity. Conditions such as tetralogy of Fallot (ToF), Alagille syndrome and Williams syndrome frequently involve branch pulmonary artery stenoses, abnormal right-ventricular (RV) loading, and pulmonary perfusion imbalance. Patients with complex cases often require catheter-based interventions or open-heart surgeries in early childhood, and clinical decision-making is further complicated by high anatomic variability, limited use of intra-operative angiograms to reduce radiation and contrast agent exposure, the permanent nature of device deployments which prevents iterative testing, and the uncertainty of compounding downstream effects in multi-staged procedures.

The rapid development of patient-specific computational hemodynamics simulations has been invaluable for making in silico predictions of post-intervention changes in key metrics such as pressure and flow split. A common transcatheter intervention is stenting, a minimally invasive procedure that enlarges the lumen of a local vessel region by expanding a cylindrical wire-mesh device. However, as stents and other interventional implants such as grafts and occluders become more routine, an important practical gap has emerged: the need to rapidly create anatomically faithful, device-induced geometric modifications on patient-specific vascular models. Existing segmentation and modeling tools excel at reconstructing anatomies in diseased states, but they often fall short in capturing realistic stent-induced shape changes: a near-cylindrical luminal expansion with axial fall-off, junction surface preservation, and secondary nearby deformations. Full solid-mechanics simulations could capture these effects, but they are often prohibitively slow for interactive treatment planning or multi-scenario testing under clinical decision-making timescales. Beyond the time considerations, full mechanics simulations can be unnecessary for the purpose of clinical planning for balloon-expandable stenting: the interventionalist targets an intended final stent diameter through balloon inflation, so the practical need is to produce post-intervention virtual geometry given pre-determined dimensions rather than to predict post-intervention dimensions.

Our study addresses this need by presenting a virtual-stenting mesh deformation method that enables controlled, anatomically plausible diameter changes along targeted vessel segments. The method is designed to integrate seamlessly with established cardiovascular modeling workflows and to operate at interactive speeds, allowing users to quickly iterate on device location and target diameters while preserving critical geometric features at bifurcations and ostia. We benchmark the approach against alternative deformation strategies and validate on real-world patient-specific CHD models, and we assess its efficacy using both geometric metrics and relevant hemodynamic quantities.

\subsection{Background}
\label{sec:background}

The advent of dedicated software packages for studying cardiovascular hemodynamics enabled an abundance of opportunities to gain insight into the flow physics of various regions of the vascular system. SimVascular~\cite{updegrove_simvascular_2017} is one such open-source package that provides an integrated workflow combining image processing of computed tomography (CT) and magnetic resonance imaging (MRI) volumetric data, geometric model smoothing and refinement, volumetric mesh generation, and computational fluid dynamics (CFD) simulation on large computing clusters. Recent works have extended prior capabilities to provide support for model manipulation. For example, svMorph~\cite{pham_svmorph_2023} and morphMan~\cite{kjeldsberg_morphman_2019} introduced geometry-based techniques to augment and morph existing models to capture common diseased anatomic features such as aneurysms and stenoses. Combined with a simulation software suite, these tools make it easier to investigate a wide range of morphological changes in blood vessels. However, there remains a need for fast and accurate methods to model geometry changes resulting from stent placement. While anatomic feature morphing tools above use geometric techniques that generate round, natural shape changes with smooth transitions, stenting produces a near-cylindrical impression on the vessel lumen with a specific diameter and length, determined by the specifications of the stent and balloon selected. Existing tools can be adapted to create elongated shapes that mimic stents but the results can be inefficient and inconsistent simply because the stent shape lies outside the intended shape space.

Surgem~\cite{pekkan_patient-specific_2008} was an early work in interactive vascular anatomic feature editing that coupled free-form surface deformation operations to intuitive, physical controllers such as magnetic trackers and robotic manipulators to enable surgeons to easily create morphological edits reflecting diseased anatomy as well as surgical or interventional modifications, including balloon angioplasty. Harvis~\cite{shi_harvis_2020} is a more recent work that provides anatomic shape editing, surgical device registration for conduits and stents, and high-performance finite element (FE) simulations through the interactive platform of virtual reality (VR). Harvis generates the post-stent vascular surface by generating a secondary shape representing the medical device and using a constructive solid geometry (CSG) package to perform 3D boolean remeshing. However, this construction confines the alteration to the stented section only and does not capture shape changes in neighboring surfaces or preserve the original mesh in nearby junctions.

In addition to methods that directly manipulate the vessel shape, another category of shape editing is physics-based deformations. Regularized Kelvinlets are closed-form solutions to isotropic linear elasticity equations with regularized loads and have been applied to model soft tissue shape changes such as in blood vessels~\cite{pham_virtual_2024, ringel_comparing_2024} with volume preservation properties. Extended position-based dynamics (XPBD) is a particle-based simulation technique from computer graphics that was adapted by Pham et al.~\cite{pham_deforming_2024} to specifically simulate vascular surfaces undergoing stenting by modeling restoration energies in the vessel wall under expansion forces exerted by the stent.

At the other end of the fidelity spectrum, 3D FE methods model the full device-tissue contact problem with constitutive laws for both the stent material and the vessel wall. Gijsen et al.~\cite{gijsen_simulation_2008} demonstrated an early patient-specific FE simulation of coronary stent deployment in a 3D-reconstructed artery, resolving strut-level wall stresses to investigate re-stenosis risk. More recently, Jiang et al.~\cite{jiang_toward_2026} established a validated framework for balloon-expandable coronary stent deployment in FEBio~\cite{maas_febio_2012}, using an elastic-plastic constitutive model for the Co-Cr stent alloy and comparing FE-predicted deformation against $\mu$CT imaging; their highest-fidelity simulation required approximately 146 hours on a 192-core high-performance computing cluster.

Beyond balloon-expandable stents, similar workflows have been developed for self-expanding transcatheter devices, as well. Zelonis et al.~\cite{zelonis_integrated_2025} recently developed an FE simulation workflow for self-expanding transcatheter pulmonary valve (TPV) deployment using the SlicerHeart~\cite{lasso_slicerheart_2022} extension for 3D Slicer~\cite{fedorov_3d_2012} in conjunction with FEBio. Their approach segments the right ventricular outflow tract (RVOT), positions commercial device models (Harmony TPV-25, Alterra) interactively in a dedicated SlicerHeart module, and exports the geometry to FEBio for staged device compression and release using a Holzapfel--Gasser--Ogden (HGO) constitutive model for the vessel wall. The workflow involves an extensive meshing pipeline of several intermediate representation formats: the RVOT surface is exported from 3D Slicer as an STL, remeshed as quad-dominant shell elements in Blender~\cite{blender_online_community_blender_2026} as a PLY file and then imported into FEBio for extrusion, and the TPV device and the compression sheath are meshed in Visualization Toolkit (VTK)~\cite{schroeder_visualization_2006}, while the compression sheath is derived from the Vascular Modeling Toolkit (VMTK)~\cite{izzo_vascular_2018} vessel centerline~\cite{antiga_image-based_2008} using least squares polynomial interpolation and a vtkTubeFilter-based 3D Slicer extension. The stent is pinned to the centerline to prevent any axial movements, and stent expansion is driven by a normal displacement boundary condition (BC) applied to the compression sheath with a contact potential. The interactive device placement along the centerline can be completed in 45 seconds, while the simulation of the device deployment takes an average of 2--5 hours on a workstation-class CPU.

These full-physics FE approaches and the fast editing methods described above serve complementary purposes, separated by what question is being asked. Physics-based modeling is needed when the deployed configuration is uncertain beforehand or when the quantity of interest is itself a constitutive quantity. Self-expanding devices are an example of the first case: while the operator selects a device size from a discrete catalog, the actual deployed configuration emerges from the equilibrium between the device's outward radial force and the resistance of the surrounding anatomy. Device design and durability assessment are examples of the second case, where strut-level wall stress and strain distributions are the primary outputs of interest that provide in silico evidence for regulatory submissions. While balloon-expandable stenting also exhibits small deviations from nominal target dimensions due to elastic recoil and foreshortening, the deployed diameter and length are nevertheless largely controllable by the interventionalist, who can compensate using strategic over-inflation or post-dilation, so a runtime mechanical simulation is not necessary when the planning question is a hemodynamic one.

A further differentiator is data availability: fully patient-specific physics requires patient-specific vessel wall material properties, which are rarely measurable in clinical practice. Geometric targeting does not require these properties at runtime, though they are implicitly encoded in the manufacturer sizing tables that determine the target dimensions. As a result, geometric targeting is more reproducible across users, since its inputs are the location and dimension of the stent only. Fast geometry editing methods are therefore well-suited to hemodynamic-planning questions where the target geometry is known and the goal is to obtain the resulting flow field quickly. The two approaches face a trade-off between fidelity and computational efficiency. Achieving end-to-end real-time interactivity while resolving contact mechanics, vessel-wall constitutive nonlinearity, and stent material plasticity simultaneously remains a challenge.

Since imaging is a precursor to geometric models as well as an essential clinical tool throughout procedures, simulating it directly is a valuable approach, as well. Barak-Corren et al.~\cite{barak-corren_virtual_2025} recently developed Virtual Cath Lab within the SlicerHeart extension for 3D Slicer, which generates simulated fluoroscopic angiography images from CT data with virtual device placement and has been applied to CHD transcatheter intervention planning such as ductus arteriosus stenting and transcatheter pulmonary valve replacement. While Virtual Cath Lab addresses the downstream challenge of visualizing planned interventions under realistic fluoroscopic conditions, it uses pre-existing device geometries and does not modify the underlying vascular anatomy, which underscores the complementary need for an upstream geometry editing tool that can produce post-intervention vessel shapes.

\begin{figure}[htbp]
  \centering
  % --- Row 1 ---
  \includegraphics[trim=0 0 0 0,clip,width=0.222\columnwidth]{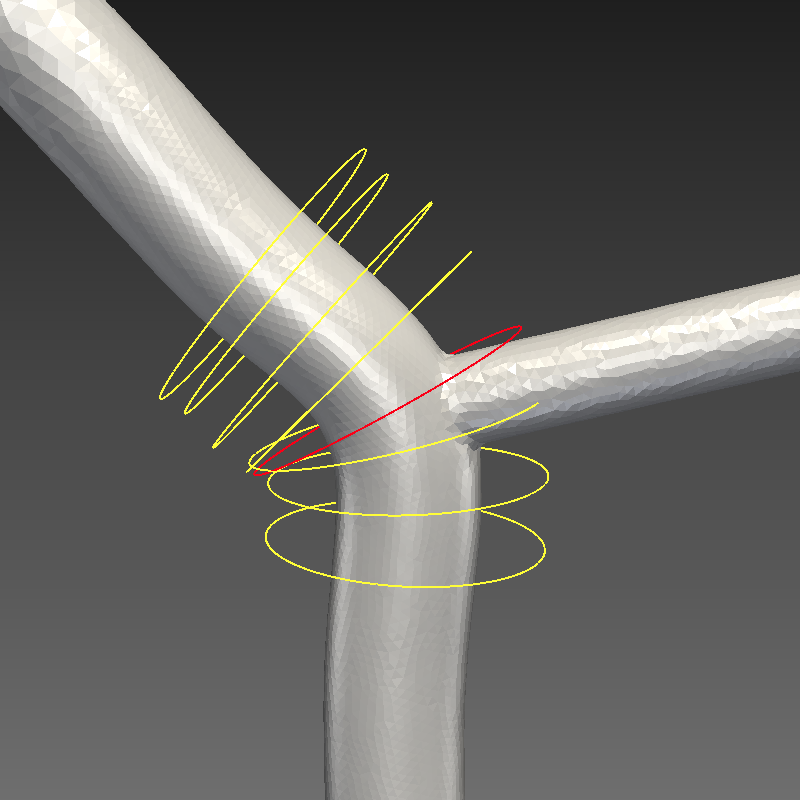}\kern-0.5pt%
  \includegraphics[trim=0 0 0 0,clip,width=0.222\columnwidth]{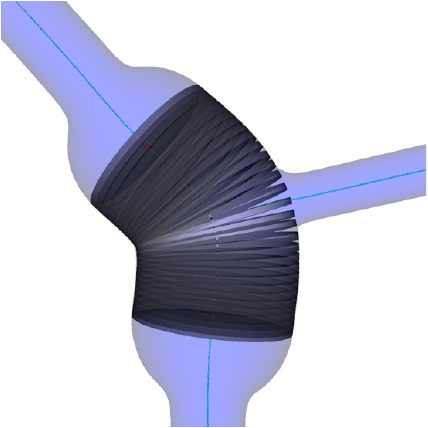}\kern-0.5pt%
  \includegraphics[trim=0 0 0 0,clip,width=0.222\columnwidth]{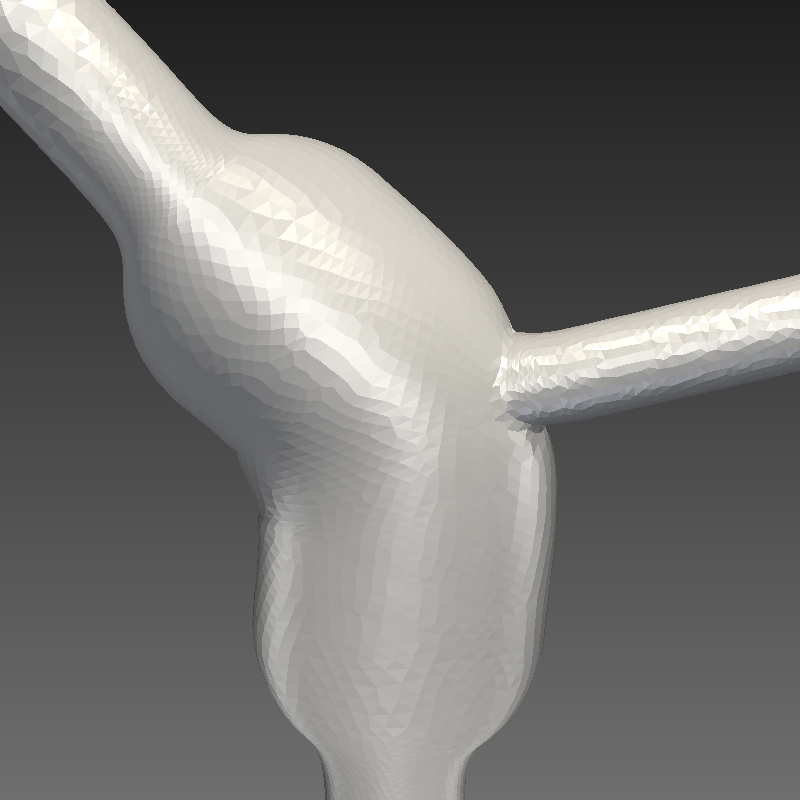}%
  % Remove the baseline glue between rows:
  \par\nointerlineskip
  % --- Row 2 ---
  \includegraphics[trim=0 0 0 0,clip,width=0.222\columnwidth]{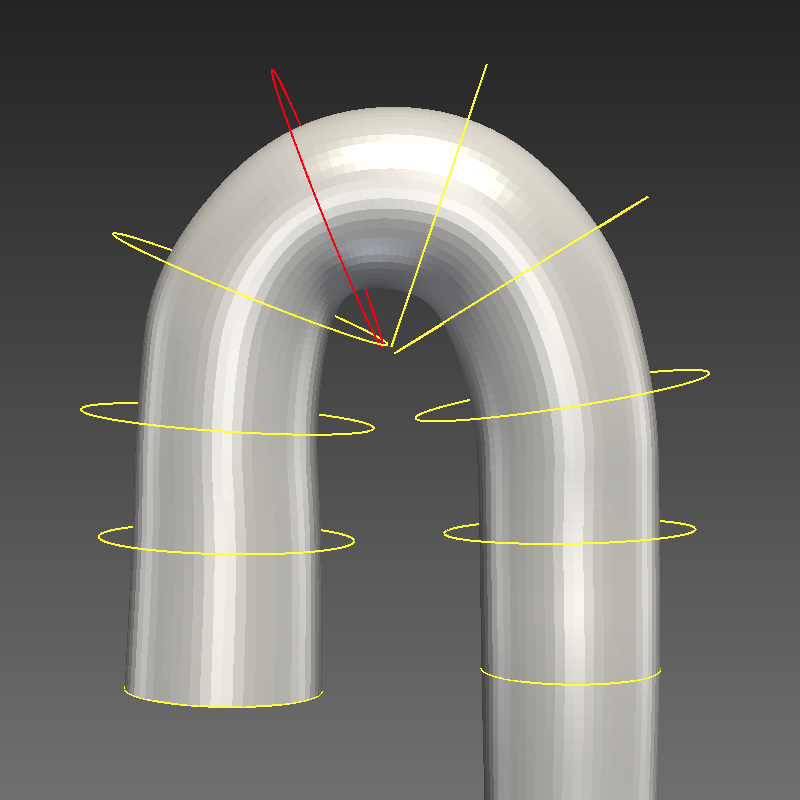}\kern-0.5pt%
  \includegraphics[trim=0 0 0 0,clip,width=0.222\columnwidth]{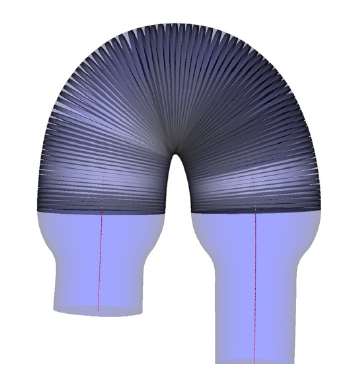}\kern-0.5pt%
  \includegraphics[trim=0 0 0 0,clip,width=0.222\columnwidth]{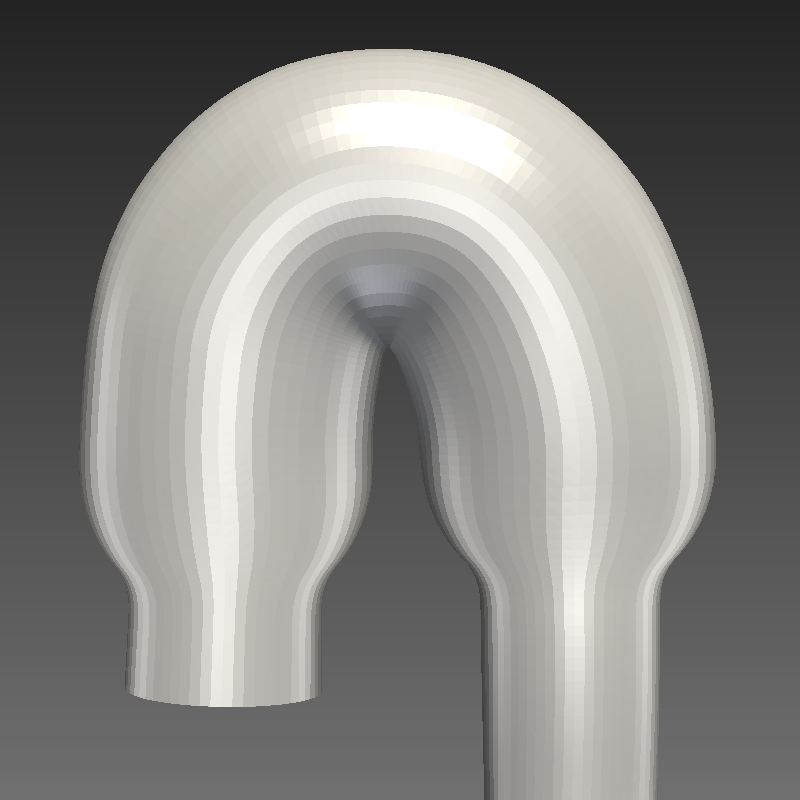}%
  \caption{Limitations of manual segmentation editing for stenting. Two synthetically constructed examples of manual segmentation editing causing contour intersections (left); screenshot of SDFStent virtually deploying the stent (middle), and the resulting surface with deployed stent produced by SDFStent (right).}
  \label{fig:synthetic-triple-examples}
\end{figure}

\begin{figure*}[htbp]
  \centering
  \includegraphics[trim=10 0 75 0,clip,width=0.333\textwidth]{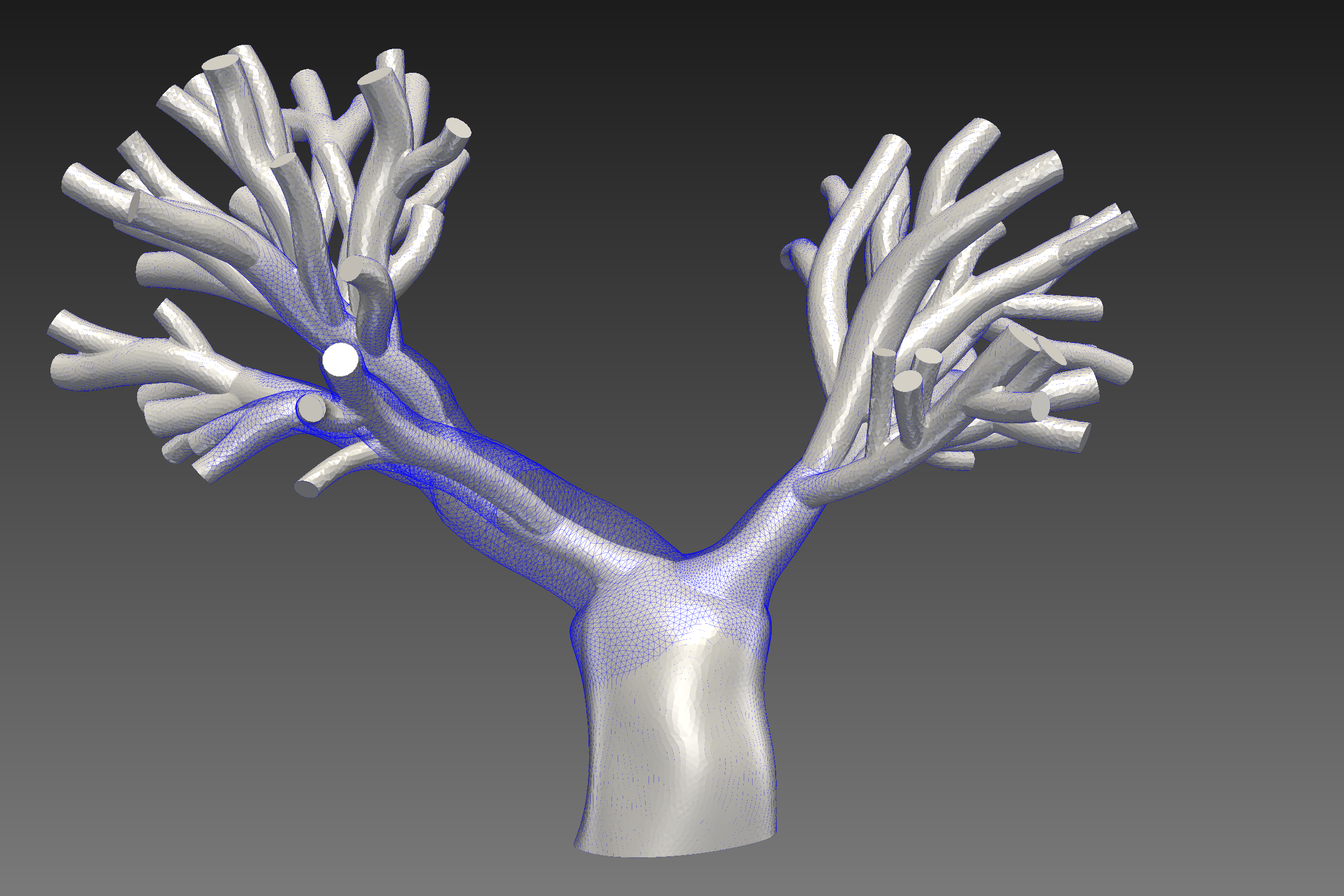}\kern-0.5pt
  \includegraphics[trim=10 0 75 0,clip,width=0.333\textwidth]{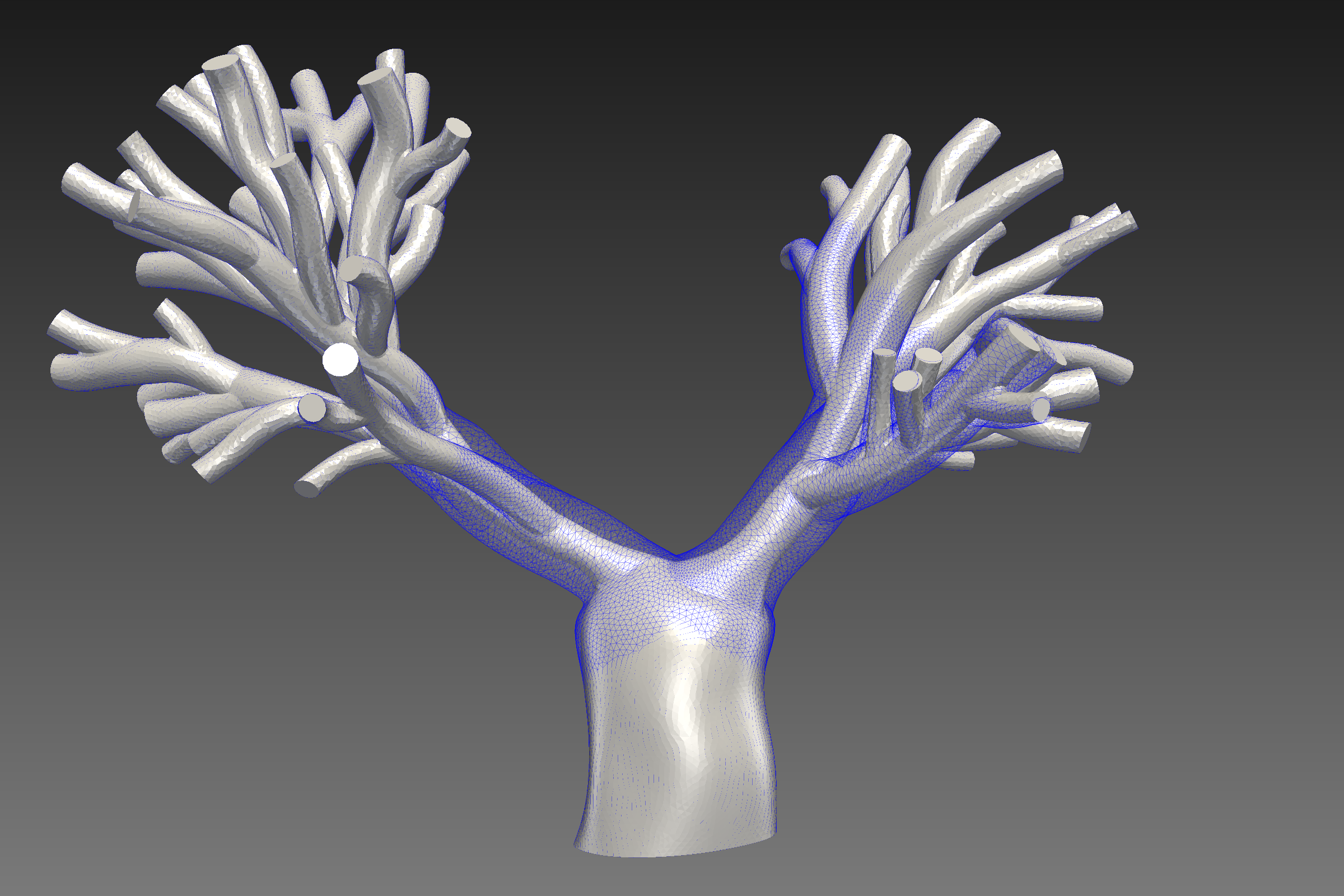}\kern-0.5pt
  \includegraphics[trim=10 0 75 0,clip,width=0.333\textwidth]{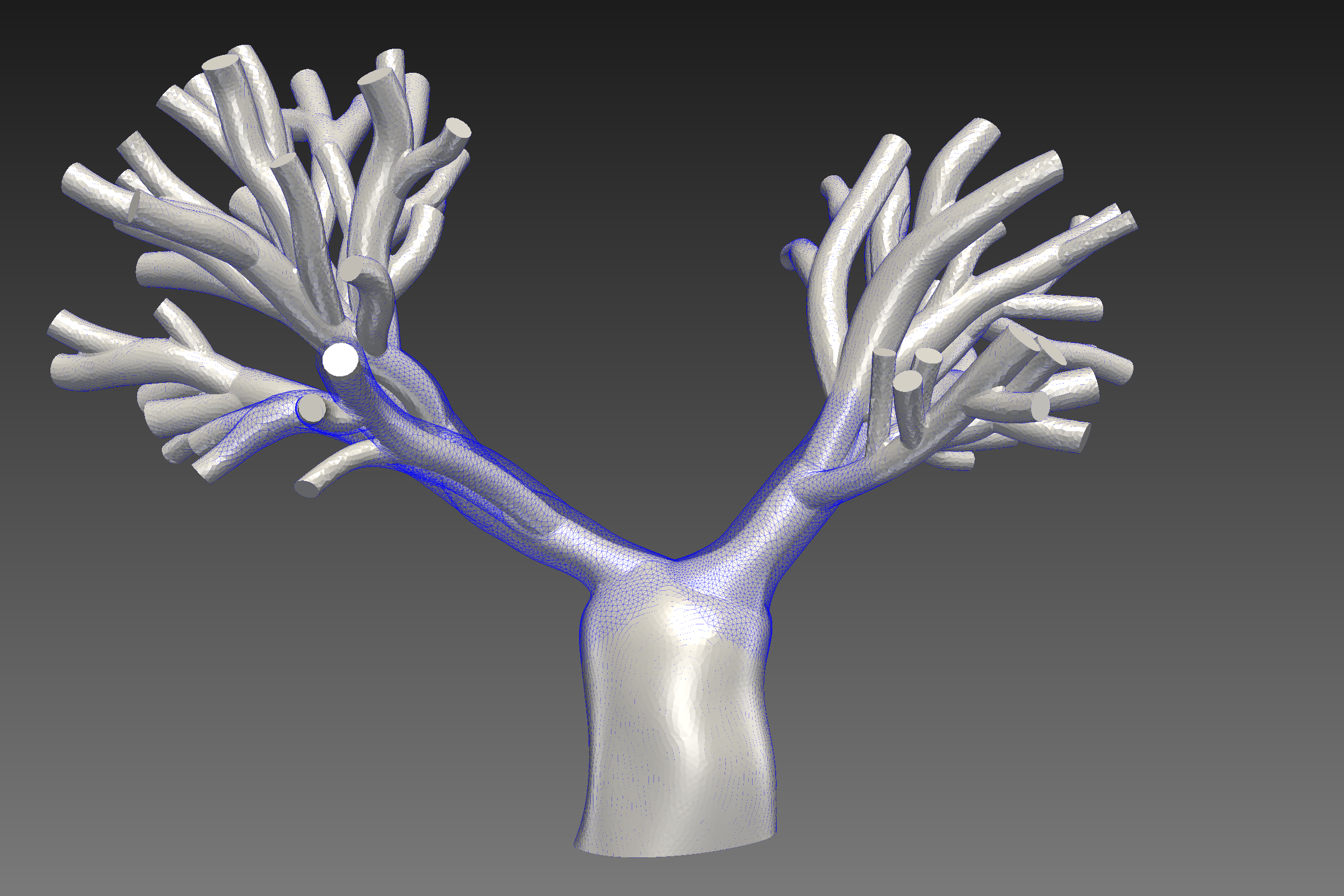}
  \caption{Visualizations of three distinct stent intervention configurations. Three sets of stent interventions are applied to the same patient-specific pulmonary artery anatomy with Alagille syndrome, targeting different combinations of main branch pulmonary arteries and daughter vessels. Each configuration was produced interactively in seconds, illustrating the rapid iteration over stent location, number, and extent enabled by SDFStent.}
  \label{fig:different-repair-configurations}
\end{figure*}

Despite the above advances in simulation and model-based editing, modeling catheter-based interventions and devices remains theoretically feasible but practically cumbersome, requiring time-consuming manual adjustments of segmentations and subsequent remeshing conversions, or high computational costs for high-fidelity simulations. Moreover, manually adjusted segmentation contours can easily intersect and fail to loft into a simple surface if the local curvature is high (Figure~\ref{fig:synthetic-triple-examples}). Lastly, reference post-operative imaging data is rarely available for many clinical scenarios, since patients often become ineligible for MRI due to the metallic material in the implanted stent, and post-operative CT is often avoided unless necessary due to radiation concerns, particularly in pediatric patients. Therefore, a strong need exists for the development of a fast, robust, and precise virtual stenting geometry editing tool to facilitate rapid 3D model modification and flexible integration with numerical simulation workflows. Figure~\ref{fig:different-repair-configurations} shows how such a tool might enable clinicians, without the assistance of geometric modeling experts, to rapidly iterate several different stent intervention configurations that can then be immediately meshed and evaluated in high-fidelity blood flow simulations.

%% --- Full-width figure: synthetic examples (row of 3 + row of 3) ---

Beyond interactive use, a computationally lightweight method also enables scripted automation: with minimal adaptation, the same deformation pipeline can be driven programmatically to sweep over stent location, length, and target diameter in a combinatorial fashion, generating large cohorts of synthetic post-operative geometries without manual intervention. Such capabilities are increasingly important for data-driven approaches to cardiovascular modeling, where reduced-order models (ROMs), surrogate-based optimization, and uncertainty quantification methods all require training datasets far larger than what manual geometry editing can feasibly produce~\cite{du_deep_2022, choi_falconbc_2026, maher_geometric_2021}.

To this end, we propose SDFStent, a signed distance function (SDF) based, volumetric sculpting technique that:

\noindent(1) Operates interactively in real time.\\
\noindent(2) Produces watertight surface meshes free of self-intersections, suitable for direct volumetric meshing.\\
\noindent(3) Extends the stent-induced deformation into a finite shell that preserves junction geometry at bifurcations and ostia.

%% --- Full-width figure: three repair configurations ---

Unlike Surgem and Harvis, we forgo developing a physical user interface with specialized hardware in favor of a graphic user interface (GUI) typical in modern 3D modeling software; however, our work shares their design philosophy of maintaining intuitive interactivity by not exposing excessive algorithmic parameters to the user. The result is a robust geometry editing tool that readily integrates stenting capabilities into many hemodynamic simulation workflows. In addition to our standalone GUI, SDFStent has been adapted into a custom module for 3D Slicer, where it coexists with diverse tools including the Virtual Cath Lab angiography simulator within the same open-source platform. This coexistence enables a natural future integration in which post-stent vascular geometries produced by our tool may be directly rendered as simulated angiograms, closing the loop between geometric planning and procedural visualization.

%% =====================================================================
%%  2.  METHOD
%% =====================================================================
\section{Methods}
\label{sec:method}

\begin{figure*}[htbp]
  \centering
  \includegraphics[width=\textwidth]{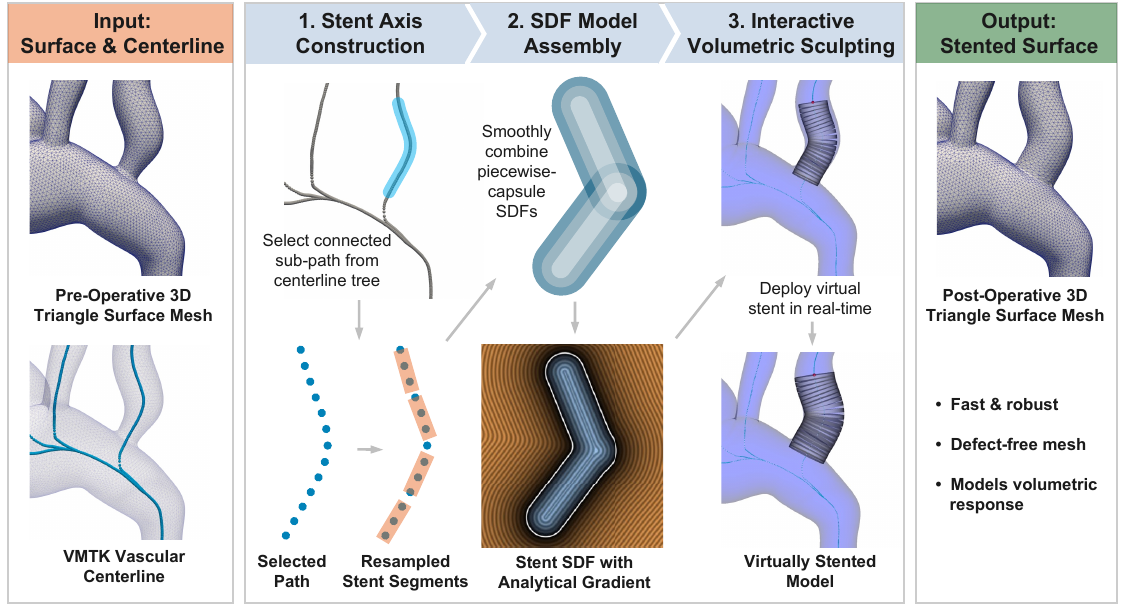}
  \caption{Overview of the virtual stenting pipeline. Given a pre-operative surface mesh and VMTK centerline as input (left), the method proceeds in three stages: (1) stent axis construction via centerline sub-path selection and segment resampling, (2) SDF model assembly by smoothly combining capsule signed distance functions piecewise, and (3) interactive volumetric sculpting to deform the vessel surface in real time. The output is a post-operative triangle surface mesh suitable for downstream CFD simulation tasks.}
  \label{fig:overall_framework}
\end{figure*}

The overall pipeline is detailed in Figure~\ref{fig:overall_framework}. SDFStent models a one-way interaction between two components: the vessel wall represented as a discrete triangle surface mesh and the stent as an SDF. During deployment of the stent, we will repeatedly query the SDF and its gradient field to compute both a magnitude and a direction of displacement at each mesh vertex in proximity to the stent. The directions of the displacement vectors are, in fact, defined independently of the mesh being deformed; instead, they are a function of the SDF only and are predetermined. On the other hand, the magnitude of the displacement vectors depends on the inter-vertex distances of the triangle mesh, which changes dynamically as the vertices move over the course of the stent deployment.

%% --- Full-width figure: overall framework ---

\subsection{Modeling the stent with SDFs}
\label{sec:modeling_stent_sdf}

Signed distance functions (SDFs) are functions that give the orthogonal distance to a surface, commonly in 2D or 3D, and they are ubiquitous in the computer graphics community. An SDF implicitly defines a surface $\partial \Omega$ as the set of points where the signed distance is exactly 0, and the distance is signed since it is positive at points outside the surface and negative at points inside. Formally we have the following:
\begin{align}
    &\phi : \mathbb{R}^3 \rightarrow \mathbb{R}, \\
    &\partial \Omega = \{x \in \mathbb{R}^3 : \phi(x) = 0\},
\end{align}

\noindent and for points outside the surface, the SDF measures the distance to the surface by the definition:
\begin{equation}
\label{eq:sdf_closed_form}
\phi(x) = d(x, \partial \Omega)\ \ \forall x \notin \Omega.
\end{equation}
Because of its simplicity and natural compatibility with distance and normal computations, SDFs are a highly useful and versatile shape representation. In our algorithm, we use SDFs to model the diverse range of large-diameter balloon-expandable arterial stents typically deployed in the pulmonary arteries and the aorta~\cite{gupta_pulmonary_2024, meadows_intermediate_2015}.

While SDFs are often stored numerically as dense 3D spatial grids, they can also be specified analytically as closed-form mathematical expressions, as in Equation~\eqref{eq:sdf_closed_form}. For our stent model, we leverage the analytical form of the ``capsule'' shape, defined as a cylinder bounded by two hemispheres of matching radii. By connecting these simple and composable capsules head-to-tail, we can adaptably approximate the curved cylindrical shapes of various physical stents. Equations~\eqref{eq:capsule_sdf_1} and~\eqref{eq:capsule_sdf_2} below define the closed-form for the capsule shape SDF. Let $a, b$ denote the endpoints of the line segment representing the axis of the capsule, let $p$ denote the point we are querying, and let $r$ denote the capsule radius, so that
\begin{align}
\label{eq:capsule_sdf_1}
    h &= \text{clamp}\Bigl(\cfrac{\overline{pa} \cdot \overline{ba}}{\norm{\overline{ba}}^2},0,1\Bigr)
\end{align}
and
\begin{align}
\label{eq:capsule_sdf_2}
    \phi_{\overline{ba}}(p) &= \norm{\overline{pa} - \overline{ba}\ h} - r \notag \\[4pt]
    &= \begin{cases}
    \norm{\overline{pa}} - r,
    & \dfrac{\overline{pa} \cdot \overline{ba}}{\norm{\overline{ba}}^2} < 0 \\[18pt]
    \norm{\overline{pa} - \overline{ba}} - r,
    & \dfrac{\overline{pa} \cdot \overline{ba}}{\norm{\overline{ba}}^2} > 1 \\[18pt]
    \norm{\overline{pa} - \textup{proj}_{\overline{ba}}(\overline{pa})} - r,
    & \text{otherwise}.
    \end{cases}
\end{align}
\noindent In short, the distance to a capsule's surface is the distance to its axis line segment offset by the radius of the capsule.
To model the shape of a fully deployed stent and subsequently its effect on the surrounding vascular geometry, we adopt an idealized model considering the deployed stent as a canal surface of constant radius or ``pipe surface''. Geometrically, a pipe surface is formed as the envelope of a family of spheres whose centers lie on a continuous curve. Letting $\Gamma : \mathbf{x} = \mathbf{c}(u) = (x(u), y(u), z(u))^{\top}$ be a regular space curve, then the resulting pipe surface can be implicitly defined by the following parameterized SDF:
\begin{align}
    f(x; u) := \norm{\mathbf{x} - \mathbf{c}(u)} - r = 0.
\end{align}
Therefore, the stent surface is uniquely parameterized by its center axis curve, and rather than computing an analytical form for the curve that represents the stent's center axis, we will approximate the center axis curve $\Gamma$ as linear splines of densely sampled points, i.e., a connected path of line segments $c_1, c_2, ..., c_m$. We can then approximate the pipe surface as the envelope of a family of capsule shapes $P_1,P_2,...,P_m$, where each capsule $P_i$ has $c_i$ as its center axis. Next, as shown in Equation~\eqref{eq:capsule_sdf_2}, for each capsule $P_i$ we can compute the corresponding SDF $\phi_{c_i}$ such that $\phi_{c_i}(p) = 0$ represents the capsule surface. Finally, we compose the $m$ capsule shapes by taking their solid union, and the resulting shape is our desired representation for the stent.

%% --- Full-width figure: naive vs smooth union comparison ---
\begin{figure}[htbp]
  \centering
  \includegraphics[trim=2 0 0 5,clip,height=0.3\columnwidth,keepaspectratio]{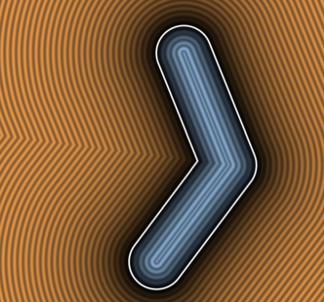}%
  \hspace{0.02\columnwidth}%
  % trim = left bottom right top
  \includegraphics[trim=0 2 0 6,clip,height=0.3\columnwidth,keepaspectratio]{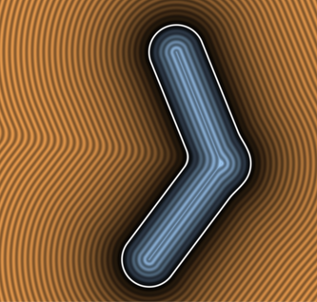}
  \caption{Comparison of the union of two 2D capsule SDFs using naive minimum (left) vs quadratic smooth minimum (right). Notice how naive minimum creates sharp creases in the field, which corresponds to $C^0$ continuity, whereas quadratic smooth minimum creates smooth bends with $C^1$ continuity. The images are created with ``Segment - distance 2D'' shader by Inigo Quilez~\cite{quilez_segment_2020} with modifications by the authors.}
  \label{fig:comparison-union-naive-vs-quadratic-smooth}
\end{figure}

\subsection{Quadratic smooth minimum of capsules}
\label{sec:smooth_min}

A primary benefit of using SDFs to represent shapes is that CSG operations become simple and elegant to perform: the union of two shapes is obtained by taking the minimum of their SDFs while the intersection of two shapes is obtained by taking their maximum:
\begin{align}
    \phi_{A\cup B}(p) &= \min\bigl(\phi_A(p), \phi_B(p)\bigr), \\
    \phi_{A\cap B}(p) &= \max\bigl(\phi_A(p), \phi_B(p)\bigr).
\end{align}
Although the resulting functions $\phi_{A\cup B}, \phi_{A\cap B}$ computed this way are not always true SDFs, in that they may violate the distance property on the interior of the shape, they always produce the correct zero level set and correct distance values on the exterior of the shape, which is sufficient for our application. Therefore, we might naively compute the SDF for $P_1\cup P_2 \cup...\cup P_m$ as
\begin{align}
\phi_{stent}(p) := \min\bigl\{\phi_{c_1}(p),...,\phi_{c_m}(p)\bigr\}.
\end{align}
However, we note that taking a naive union on the $m$ capsule shapes $P_1,P_2,...,P_m$ would result in sharp, $C^0$ surface transitions between consecutive straight sections of the capsule segments (Figure~\ref{fig:comparison-union-naive-vs-quadratic-smooth}). This is a natural consequence of using linear splines which reduce computational complexity instead of those with higher order smoothness such as cubic splines. A sharp transition in the surface of the stent shape would cause discontinuities in the displacement field in our deformation algorithm, which in turn can easily cause mesh vertices to cross over each other, creating self-intersections. To mitigate this issue, we use a smooth union operator called quadratic smooth minimum~\cite{quilez_smooth_2013}, which smoothly welds the capsules together at their interface instead of bluntly superimposing them, creating smooth $C^1$ transitions. Given a smoothing parameter $k \geq 0$, the quadratic smooth minimum is defined as follows:
\begin{equation}
\textup{smin}(\aaa,\bbb) = \begin{cases}
\min(\aaa,\bbb),  & \lvert \aaa - \bbb\rvert > k\\[6pt]
\min(\aaa,\bbb)\;-\;\displaystyle\frac{k}{4}
\Bigl(1 - \frac{\lvert \aaa - \bbb\rvert}{k}\Bigr)^{2},
& \lvert \aaa - \bbb\rvert \leq k.
\end{cases}
\label{eq:smin_form}
\end{equation}

\noindent Essentially, at locations in space that are almost equally distant to the two surfaces we are combining, we artificially pull the surface closer to that location by subtracting a quadratic offset term from the surface distance. To build intuition, consider a point lying exactly on surface B ($\bbb = 0$) and at distance $k$ from surface $A$ ($\aaa = k$). At this point, $\textup{smin}(\aaa,\bbb) = 0$, so the point lies on the smooth-union surface as well. Now perturb slightly toward $A$: $\alpha$ slightly decreases below $k$ and $\beta$ slightly increases above 0, but $\textup{smin}(\aaa,\bbb)$ still evaluates to 0, meaning the point continues to lie on the smooth-union surface even though it is now outside surface $B$ ($\beta > 0$). Another mathematically equivalent way of characterizing the smooth union is the following:

\begin{align}
\label{eq:kernel_form}
\textup{smin}(\aaa,\bbb) &= \bbb - k\,g\Bigl(\frac{\bbb-\aaa}{k}\Bigr)\\[8pt]
g(x)&=
    \begin{cases}
    0, & x < -1\\[6pt]
    x, & x > 1\\[8pt]
    \cfrac{x(x+2) + 1}{4},
    & -1 \leq x \leq 1.
    \end{cases}
\end{align}
It is clear from this formulation that the kernel $g(\cdot)$ acts as a smooth activation function interpolating between $y=0$ and $y=x$ between -1 to 1, and its derivative is continuous at both endpoints. Moreover, if we take $g(\cdot)$ to be $\max(0, x)$, commonly known as the ReLU function, then the resulting function becomes the ordinary $\min$ function.

\begin{figure*}[htbp]
  \centering
  \includegraphics[width=1.0\textwidth]{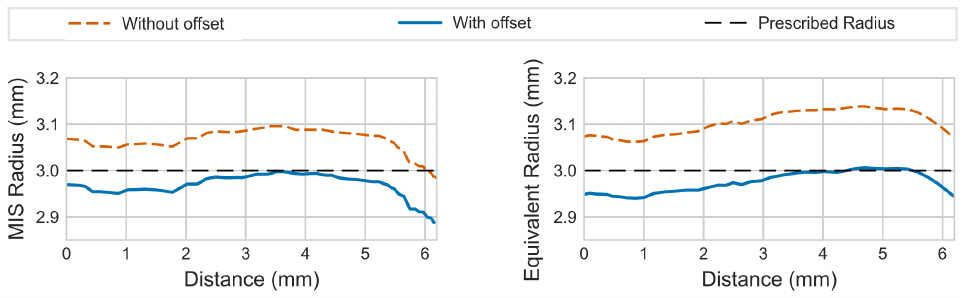}%
  \caption{Maximum inscribed sphere (MIS) radius (left) and equivalent radius (right) along a representative virtually stented vessel, with and without the radius offset correction. The equivalent radius is defined as $r_\mathrm{eq} = \sqrt{A/\pi}$, where $A$ is the cross-sectional area. The dashed line indicates the prescribed stent radius of 3.0~mm.}
  \label{fig:inflation-fix-comparisons}
\end{figure*}

Figure~\ref{fig:comparison-union-naive-vs-quadratic-smooth} shows another important consequence of applying the smooth min function to our stent shape: the ``elbow'' of the combined capsule is slightly inflated. This can be explained by substituting $\aaa = \bbb$ in Equation \eqref{eq:smin_form}:
\begin{align}
\textup{smin}(\aaa,\bbb) = \aaa - \cfrac{k}{4}. 
\end{align}
In regions of space where the two capsule shapes are superimposed, the resulting SDF value is reduced by exactly $\frac{k}{4}\textup{mm}$, leading to a slight inflation of the shape by $\frac{k}{4}\textup{mm}$. As the stent model inflates and its radius increases, the overlap between segments grows as well. And, since our stent shape is densely segmented, the capsule joints are also numerous and dense, so the inflation effect becomes sufficiently prominent that the entire stent slightly inflates by up to $\frac{k}{4}\textup{mm}$. Therefore, to correct for this difference between nominal prescribed radius and actual radius, we decrease the prescribed radius of the stent by $\frac{k}{4}\textup{mm}$. Figure~\ref{fig:inflation-fix-comparisons} shows the correction effect in a representative virtual stent deployment. With the offset term, SDFStent produces a stented vessel whose maximum inscribed sphere (MIS) radius and equivalent radius are both upper bounded by the prescribed radius. Without the offset, the resulting MIS radius overshoots the prescribed radius at its peak by exactly $\frac{k}{4}\textup{mm}$ ($k = 0.4\textup{mm}$ in this example), and the equivalent radius overshoots by even more.

%% --- Full-width figure: offset term comparison ---
 
Lastly, note that the quadratic smooth minimum is not strictly associative, so we adopt a straightforward ordering convention for the stent construction: apply the operator pairwise, from the most distal to the most proximal point of the stent axis. In practice, the ordering has negligible effect on the resulting surface, since the capsule joints are connected in a chain, and the adjacent smin transition zones at consecutive joints are well-separated relative to our chosen smoothing scale $k$ (Section~\ref{sec:free_params}). Thus, the stent SDF is defined as the following nested composition:
\begin{align}
\phi_{stent}(p) := \textup{smin}\biggl( \dots \textup{smin}\Bigl( \textup{smin}\bigl(\phi_{c_1}(p),\phi_{c_2}(p)\bigr), \phi_{c_3}(p) \Bigr) \dots , \phi_{c_m}(p) \biggr)
\end{align}
where $c_1,...,c_m$ is the sequence of line segments that make up the stent axis, defined below.

\subsection{Constructing the stent axis}
\label{sec:stent_axis}

Following the technique adopted in~\cite{pham_virtual_2024, pham_deforming_2024}, we use the vascular centerline object as the axis of our SDF stent model. This vascular centerline can be computed by the VMTK library using the triangle surface mesh as input, and it encapsulates quantitative information about the surface mesh geometry such as MIS radius and vascular tree topology. Formally, the centerline object $\mathcal{C}$ is a family of paths $C_i$, each of which is a sequence of connected line segments $c^{(i)}_k$. Each path $C_i$ represents a continuous run along a particular branch in the vascular tree, and each point is the center of a maximally inscribed sphere at that location:
\begin{align}
    \mathcal{C} &= C_1 \cup C_2 \cup ... \cup C_n, \\
    C_i &= c^{(1)}_i \cup c^{(2)}_i \cup ... \cup c^{(m_i)}_{i}.
\end{align}
Since a physical stent is always deployed inside a contiguous region of the vascular tree but can span across segments of parent and daughter branches, we additionally organize $C_i$ into a hierarchy by computing an ancestry map based on the centerline tree topology.

The stent axis object is a connected path $\mathcal{S}$ that is a subset of the vascular centerline. For instance, in Figure~\ref{fig:stent-axis-sampling} below, $\mathcal{S}$ is a subset of $C_2 \cup C_3$. Since the stent axis must be piecewise-linear to parametrize the piecewise-capsule SDF shape, we resample $\mathcal{S}$ at a fixed arc-length interval determined by the desired capsule length. The result mimics the uniformly segmental structure of modern commercial stents~\cite{pan_structural_2021}, and decouples the stent construction from the input centerline point density or representation. 

%% --- Full-width figure: stent axis sampling ---
\begin{figure}[htbp]
  \centering
  \includegraphics[trim=10 10 15 10,clip,width=0.5\columnwidth]{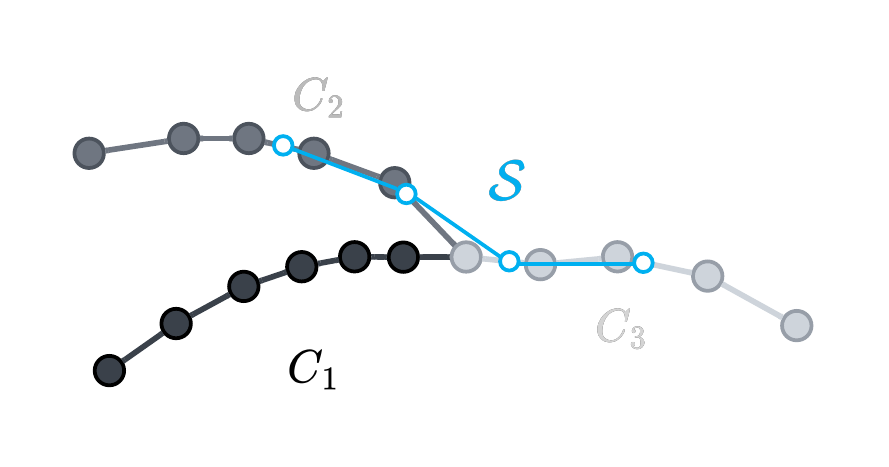}
  \caption{Example construction of the stent axis $\mathcal{S}$ (cyan) from a vascular centerline object composed of three paths $C_1$, $C_2$, $C_3$. The stent axis spans a sub-path of $C_2 \cup C_3$ across the junction and is resampled at equal arc-length intervals (open circles) to produce the piecewise-linear segments used as capsule axes in the SDF model. Filled circles denote the original centerline points.}
  \label{fig:stent-axis-sampling}
\end{figure}

\subsection{Deformation field from SDF gradient}
\label{sec:deformation_field}

From elementary vector calculus, the gradient of the SDF is orthogonal to its level sets everywhere. To expand our virtual stent, our algorithm simply steps through the level sets of the stent SDF following its gradient, letting the stent surface grow uniformly radially outward while remaining stationary along the centerline, without axial sliding (a similar no-sliding constraint is enforced in~\cite{zelonis_integrated_2025} via a zero-displacement BC). By assigning the SDF gradient itself as the direction of our mesh deformation, we naturally align the movement of the mesh with the expansion of the stent surface, ensuring that the deforming mesh stays conformed to the stent geometry without any tangential movements. This is a purely geometric construction: it makes no claims about the contact mechanics of stent deployment, which would depend on stent cell design, tissue properties, and deployment technique. Rather, it provides a simple and general surface edit that is agnostic to the specific mechanical interaction between the stent and the vessel wall.

To compute the gradient field of the SDF, we start with Equation~\eqref{eq:kernel_form} and apply the chain rule:
\begin{align}
    f(\aaa,\bbb)&= \bbb - k\,g\Bigl(\frac{\bbb-\aaa}{k}\Bigr), \\
    \nabla f(\aaa,\bbb) &= g'\Bigl(\frac{\bbb-\aaa}{k}\Bigr)\nabla \aaa + \biggl(1-g'\Bigl(\frac{\bbb-\aaa}{k}\Bigr)\biggr)\nabla \bbb, \\[4pt]
    &=
         \begin{cases}
        \nabla\,\bbb, & \cfrac{\bbb-\aaa}{k} < -1\\[4pt]
        \nabla\,\aaa, & \cfrac{\bbb-\aaa}{k} > 1\\[8pt]
        \cfrac{1}{2}\biggl(\Bigl(\cfrac{\bbb-\aaa}{k}+1\Bigr)\nabla \aaa + \Bigl(1-\cfrac{\bbb-\aaa}{k}\Bigr)\nabla \bbb\biggr),
        & -1 \leq \cfrac{\bbb-\aaa}{k} \leq 1.
         \end{cases}
         \label{eq:sdf_analytical_gradient}
\end{align}
If we suppose capsule $\aaa$ has line segment $\overline{ba}$ as the axis and $\bbb$ has $\overline{dc}$ as the axis, then

\begin{equation}
\label{eq:alpha_beta_gradient}
\nabla \aaa = \nabla \phi_{\overline{ba}}(\cdot\;; r),\; \nabla \bbb = \nabla \phi_{\overline{dc}}(\cdot\;; r),
\end{equation}
and it is straightforward to compute the analytical gradient of Equation~\eqref{eq:capsule_sdf_1} and~\eqref{eq:capsule_sdf_2} to obtain the gradient field of each individual capsule, so that
\begin{align}
&\nabla \phi_{\overline{ba}}(p\;; r) = \notag \\[4pt]
&\begin{cases}
    \dfrac{\overline{pa}}{\norm{\overline{pa}}}=\dfrac{p-a}{\norm{p-a}},
    & \dfrac{\overline{pa} \cdot \overline{ba}}{\norm{\overline{ba}}^2} < 0
    \\[14pt]
    \dfrac{\overline{pa}-\overline{ba}}{\norm{\overline{pa}-\overline{ba}}}=\dfrac{p-b}{\norm{p-b}},
    & \dfrac{\overline{pa} \cdot \overline{ba}}{\norm{\overline{ba}}^2} > 1
    \\[14pt]
    \dfrac{\overline{pa} - \textup{proj}_{\overline{ba}}(\overline{pa})}
          {\norm{\overline{pa} - \textup{proj}_{\overline{ba}}(\overline{pa})}} =
    \dfrac{(I-\frac{\overline{ba}\,\overline{ba}^{\top}}{\norm{\overline{ba}}^2})\overline{pa}}
          {\norm{(I-\frac{\overline{ba}\,\overline{ba}^{\top}}{\norm{\overline{ba}}^2})\overline{pa}}},
    & 0 \leq \dfrac{\overline{pa} \cdot \overline{ba}}{\norm{\overline{ba}}^2} \leq 1.
\end{cases}
\label{eq:sdf_stent_analytical}
\end{align}
Lastly, substituting Equations~\eqref{eq:sdf_stent_analytical} and~\eqref{eq:alpha_beta_gradient} into~\eqref{eq:sdf_analytical_gradient} gives the full gradient field of our stent SDF, which can be understood as a continuous piecewise function composed of clamped linear interpolations of the surface normals of the stent capsules. The computation consists entirely of vector addition, linear projection and linear interpolation, which makes it highly parallelizable and efficient to compute. The presence of linear interpolation here is a direct benefit of our choice of quadratic smooth minimum over other forms such as cubic or exponential.

\subsection{Computing final displacements}
\label{sec:final_displacements}

Now we are ready to describe the full algorithm. By choosing to parameterize the stent with an axis curve and a radial offset, our construction naturally defines a family of pipe surfaces that captures the full range of inflated states of a stent, from its fully crimped initial state where the radius is minimal to the final expanded state where the prescribed radius is reached. This means only a single parameter $r$ is needed to interpolate between the initial shape and the final shape of the stent while preserving the fully analytical form of the SDF at any intermediate state:
\begin{align}
    &\textup{for each } r \in [r_{0}, r_{t}], \notag \\
    &\phi_{c_i}(p; r) := \norm{\overline{pa} - c_i\ h} - r, \\
    &\phi_{stent}(p; r) := \textup{smin}\Bigl(...\textup{smin}\bigl(\phi_{c_1}(p; r),\phi_{c_2}(p; r)\bigr)...,\phi_{c_m}(p; r)\Bigr).
\end{align}
To summarize the deformation algorithm: we initialize the stent at the prescribed location along the centerline with radius $r_0$ , then gradually increase the radius by $\Delta r$ until it reaches the desired specification $r_t$ while imposing displacement vectors on the vertices in contact with the stent and the vertices close to those in contact. Algorithm 1 below provides the pseudo-code for the displacements computation:

\begin{algorithm}[H]
\caption{\textsc{compute\_displacements}}\label{alg:sdf_displacements}
    \begin{algorithmic}[1]
    \Require The stent's SDF $\phi_{stent}(\cdot\;; r)$, contact threshold $d_{\mathrm{con}}$, influence radius $d_{\mathrm{infl}}$
    \For{$r$ in range $(r_{\mathrm{init}}, r_{\mathrm{target}}, \Delta r)$}
      \State $\mathcal{V}_{\mathrm{con}}\gets \{\,v\in \mathcal{V} \mid \phi_{stent}(v; r_{\mathrm{cur}})<d_{\mathrm{con}}\,\}$
      \State $\mathcal{V}_{\mathrm{infl}}\gets \{\,w\in \mathcal{V} \mid \exists v\in\mathcal{V}_{\mathrm{con}} :\;\|w - v\|<d_{\mathrm{infl}}\,\}$
      \State $\mathbf{u}\gets \textsc{get\_base\_displacements}(\mathcal{V}_{\mathrm{infl}},\,\phi_{stent},\,d_{\mathrm{infl}})$
      \State $\boldsymbol{\alpha}\gets \textsc{get\_alphas}(\mathcal{V}_{\mathrm{infl}},\,\mathcal{V}_{\mathrm{con}},\,d_{\mathrm{infl}})$
      \State $\mathbf{u}\;\gets\;\mathbf{u}\odot\boldsymbol{\alpha}$
      \State $\mathbf{\mathcal{V}}\;\gets\;\mathbf{\mathcal{V}}+\mathbf{u}$
    \EndFor
    \end{algorithmic}
\end{algorithm}

The $\textsc{get\_base\_displacements}$ function is outlined in Algorithm 2. $\textsc{get\_base\_displacements}$ computes the analytical gradient of the SDF for the stent and normalizes it, then scales it by the radius increment $\Delta r$ and multiplies it by the following distance-based fall-off function with compact support:
 
\begin{align}
\label{eq:fall_off}
\textsc{fall\_off}(d) = \begin{cases}
\displaystyle\biggl(1 - \Bigl(\frac{d}{d_{\mathrm{infl}}}\Bigr)^2\biggr)^2,
 & 0 \le d \le d_{\mathrm{infl}}\\[8pt]
0,  & \text{otherwise}.
\end{cases}
\end{align}

\begin{algorithm}[H]
\caption{\textsc{get\_base\_displacements}}\label{alg:mesh_deformation}
    \begin{algorithmic}[1]
    \Require Vertices being influenced $\mathcal{V}_{\mathrm{infl}}$, the stent's SDF $\phi_{stent}(\cdot\;; r)$, influence radius $d_{\mathrm{infl}}$, radius increment $\Delta r$
      \State $\mathbf{u}\;\gets\;\nabla\,\phi_{stent}(\mathcal{V}_{\mathrm{infl}}\;; r)\; /\;
      \norm{\nabla\,\phi_{stent}(\mathcal{V}_{\mathrm{infl}}\;; r)}$
      \State $\mathbf{u}\;\gets\;\Delta r \,\cdot\, \textsc{fall\_off}\bigl(\phi_{stent}(\mathcal{V}_{\mathrm{infl}}\;; r)\bigr)\;\odot\;\mathbf{u}$ \\
      \Return $\mathbf{u}$
    \end{algorithmic}
\end{algorithm}

The prefactor $\Delta r$ sets the step magnitude so that a vertex in contact (where $\phi_{stent} \approx 0$ and $\textsc{fall\_off} = 1$) is displaced outward by exactly $\Delta r$, matching the radial advancement of the stent SDF level set per iteration. Equation~\eqref{eq:fall_off} then confines the rest of the deformation to a finite region surrounding the stent, smoothly and monotonically interpolating from 1 to 0. We base our fall-off kernel on the one proposed in~\cite{gain_warp_2005} because of its single free parameter and low derivative magnitude, which permit large, fold-over-free deformation steps as vertices accumulate displacements across iterations. Since the fall-off function is independent of the stent shape, other profiles can be readily adopted. Note that because $\mathcal{V}_{\mathrm{infl}}$ is constructed as the discrete $d_{\mathrm{infl}}$-neighborhood of $\mathcal{V}_{\mathrm{con}}$, large discrete increments in the stent expansion can create continuous but sharp transitions at the two axial ends of the stent, which compound over repeated deformation steps into visible ``staircase'' artifacts. To suppress this, we apply $\textsc{get\_alphas}$ (Equation~\eqref{eq:get_alphas}), which is a linear blending mask that tapers each vertex's displacement to zero at the boundary of $\mathcal{V}_{\mathrm{infl}}$. We tested linear, quadratic, and exponential blending and found all three eliminated staircase artifacts equivalently across our test cases; therefore, linear blending was retained as the simplest of the three.

\begin{align}
\label{eq:get_alphas}
\textsc{get\_alphas}(p,q) = \begin{cases}
\displaystyle 1 - \frac{\norm{p-q}}{d_{\mathrm{infl}}},
 & \norm{p-q} \le d_{\mathrm{infl}}\\[8pt]
0, & \text{otherwise}.
\end{cases}
\end{align}

%% =====================================================================
%%  3.  RESULTS
%% =====================================================================
\section{Results}
\label{sec:results}
Throughout this section, dimensional quantities of stents and vessels are reported as diameters rather than radii, in line with clinical convention; this differs from the radius-based parameterization used in the Methods section, which aligns with the SDF capsule formulation.

With a goal to maximize ease of use and minimize requirements on systems and hardware, we implemented SDFStent as a GUI application in Python using PyQt6 and the well-supported accelerator libraries NumPy, SciPy, and JAX. As a common standard in computational biomedicine, we adopt the Visualization Toolkit Polygonal Data (VTP) format from the VTK library for the input and output files.

Both our custom GUI application~\footnote{\url{https://github.com/SimVascular/svMorph} (accessed 13 May 2026)} with scriptable application programming interface (API) and the 3D Slicer module~\footnote{\url{https://github.com/SimVascular/SlicerSimVascular} (accessed 13 May 2026)} are open-source under a permissive license and published on GitHub. The 3D Slicer module is available in the Extension Manager under the SimVascular extension. We developed our code on a desktop machine running Ubuntu 20.04 with an Intel Core i7-8700 @ 3.20GHz CPU and additionally tested on five Mac computers with Apple silicon and two Windows computers released within the past six years. Our GUI application does not require a discrete GPU, and it runs in real time (50--200 frames per second) on most hardware released in the past ten years. Apart from screenshots, all visualizations and figures were created using SimVascular, ParaView~\cite{ahrens_paraview_2005} and Blender.

\begin{figure*}[htbp]
  \centering
  \includegraphics[width=\textwidth]{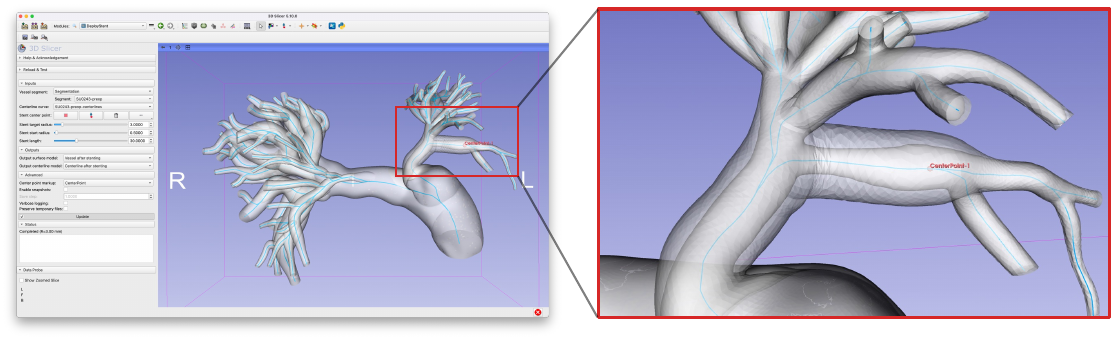}
  \caption{Virtual stent deployment applied to the pulmonary artery model of Patient B using our SDFStent module in 3D Slicer. The pre-operative and post-operative vessel geometries (white) are shown overlaid on the vessel centerline (teal), with the module control panel to the left. The red box marks the magnified region shown on the right, centered around the junction-crossing stenosis in the LPA that received virtual stenting in our testing.}
  \label{fig:slicer-demo-SU0243}
\end{figure*}

Figure~\ref{fig:slicer-demo-SU0243} showcases our real-time interactive application producing a variety of geometry edits which represent pulmonary artery stenting on a patient-specific anatomic model from a patient with Alagille syndrome. The model depicted is Patient B from the study by Yang et al.~\cite{yang_adaptive_2016} on post-virtual-repair predictions of peripheral pulmonary artery stenosis.

%% --- Full-width figure: SU0243 results ---

\subsection{Comparison to prior works}
\label{sec:comparison_prior_works}

We compared SDFStent against three prior works in patient-specific anatomic model editing with publicly available code bases, which we will briefly introduce:

\noindent(1) XPBD is a physics-based, particle simulation technique that was applied by Pham et al.~\cite{pham_deforming_2024} to model the deployment of stents.

\noindent(2) Regularized Kelvinlets based shape editor (RK)~\cite{pham_virtual_2024} is a real-time volumetric sculpting algorithm based on analytical solutions to the homogeneous linear elasticity equation.

\noindent(3) morphMan~\cite{kjeldsberg_morphman_2019} is a suite of vascular morphology manipulator tools implemented using the Voronoi implicit representation of tubular surfaces.

Of these, XPBD is the only method explicitly designed for virtual stent deployments and therefore serves as our primary comparator. The morphMan tool suite is designed to manipulate cross-sectional area and explicitly mentions removal of stenoses, which matches the intention of stenting, while RK is a geometric feature editor built for adding vascular lesion features such as aneurysms and stenoses to existing models. Since publicly available geometry editing applications specifically dedicated to stenting are exceedingly rare, we repurposed both for inclusion. Namely, we invoked morphMan twice to combine two adjacent regions of interest into a single stented region, and we composed multiple virtual aneurysms in RK to mimic the tubular shapes of stented vessels.

We tested all four methods on the pulmonary artery model for Patient B shown in Figure~\ref{fig:slicer-demo-SU0243}, where a stenosis $40\%$ in minimum luminal area (relative to the average area of the two distal ends) is located. The task was to deploy a 6.0~mm diameter stent. This is straightforward for both SDFStent and XPBD as the stent diameter is an input parameter for both. For RK, we adapted the elongated aneurysm function of the tool, which takes starting and ending points along the centerline and inflates the vascular model along the specified region where the cross-sectional area of the midpoint of the selected region is increased by a specified percentage. Based on a pre-stent midpoint cross-sectional area of 3.98~mm$^2$ and a target of 6.0~mm in diameter, we use $710\%$ as the input. Finally, for morphMan, the ``manipulate area'' subclass was chosen for our purposes as it uniformly scales up the cross-sectional area of a region.

We had limited flexibility over the use of morphMan since the tool only allows either specifying two points along a contiguous section of the vessel or automatically detecting the region between the inlet and the first bifurcation. While morphMan does offer another interpolation-based editing method under the ``stenosis'' subclass that can remove a stenosis by linearly interpolating the cross-sectional area between two parts of a vessel segment, this method does not handle stenting that crosses over a junction, and also cannot remove diffuse stenoses that do not have a focal ``hourglass'' shape. Since the stent deployment region in our test case is extensive and spans across the junction between the left pulmonary artery (LPA) and its first daughter vessel, we chose to apply the same area manipulation method twice, with the diameter increase percentage specified as $79.74\%$ and $101.68\%$ respectively so that the two ends of the stented region in the resulting surface closely reached the prescribed diameter of 6.0~mm.

\begin{figure}[htbp]
  \centering
  \includegraphics[width=0.55\columnwidth]{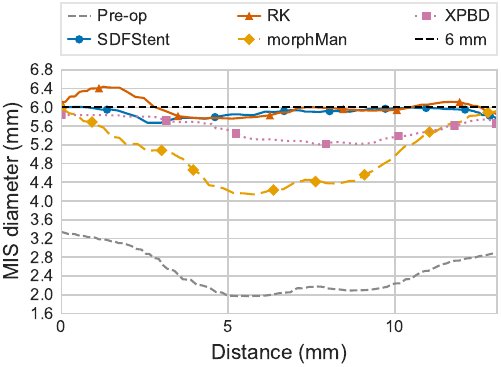}%
  \caption{Plot of the MIS diameters of the virtually stented vascular model of Patient B after applying four distinct methods. We started with the same input mesh and centerline, and the same contiguous section consisting of the LPA and its first daughter vessel was selected and the method of choice for editing the shape was applied. The MIS diameter along the same stented region of each edited model was computed and plotted.}
  \label{fig:6mm-comparison-between-methods}
\end{figure}

Figure~\ref{fig:6mm-comparison-between-methods} compares the MIS diameter along the virtually stented section of the vascular model for Patient B. The three methods that produced diameter distributions closest to the 6.0~mm target are SDFStent, XPBD, and RK. RK resulted in sections that grew beyond the prescribed target, whereas SDFStent and XPBD both ensured that the entire post-stent section had cross-sectional diameters $\leq$ 6.0~mm. XPBD undershoots the prescribed diameter by roughly 9\% in the mean (5.48 versus 6.00~mm) due to its physics-based elastic membrane model, while SDFStent matches the prescribed diameter tightly with small variance (5.92 $\pm$ 0.08~mm) since the SDF level set is defined by the target diameter directly. Notably, the diameter distribution produced by XPBD aligns with the 7\%--10\% elastic recoil typically observed in clinical stent deployments~\cite{aziz_stent_2007, carrozza_vivo_1999}, whereas SDFStent does not encode recoil behavior. Our method's strict adherence to the prescribed diameter and length instead allows one to incorporate clinically observed effects such as recoil or intentional over-inflation by directly prescribing the adjusted post-deployment diameter as input.

%% --- Full-width figure: method comparison radii ---

Table~\ref{table:prior-works-results-comparison} compares the four methods. Reported times are wall-clock end-to-end durations on identical hardware, measured from invocation of the deformation routine to its return for a single stent deployment on the test geometry; they exclude operator interaction and pre-processing steps such as centerline extraction. Times are single-run measurements and are intended to demonstrate order-of-magnitude separation between methods rather than precise benchmarking. Against XPBD, the only stenting-specific comparator, SDFStent is approximately 100$\times$ faster (1.5~s versus 160.5~s), making interactive deployment practical without sacrificing geometric fidelity in the stented region. RK, while faster than XPBD, is nonetheless one to two orders of magnitude slower than our method.

\begin{table}[htbp]
  \centering
  \begin{threeparttable}
  \caption{Summary of the virtual stenting results obtained using SDFStent, XPBD, RK, and morphMan.}
  \label{table:prior-works-results-comparison}
  \small
  \setlength{\tabcolsep}{3pt}
  \begin{tabular}{l|rrrr}
    \toprule
    Result    & Min (mm) & Max (mm) & Mean $\pm$ SD (mm) & Time (s) \\
    \midrule
    Pre-op & 1.96 & 3.34 & 2.56 $\pm$ 0.44 & -- \\
    SDFStent    & 5.68 & 6.00 & 5.92 $\pm$ 0.08 & 1.5 \\
    XPBD    & 4.80 & 5.84 & 5.48 $\pm$ 0.26 & 160.5 \\
    RK    & 5.40 & 6.44 & 6.00 $\pm$ 0.22 & 84.0 \\
    morphMan    & 4.14 & 5.98 & 5.06 $\pm$ 0.60 & 861.0 \\
    \bottomrule
  \end{tabular}
  \begin{tablenotes}[flushleft]
    \footnotesize
    \item All dimensional values reported as diameters. Times are wall-clock end-to-end measurements for a single stent deployment on the same test geometry, on identical hardware. The morphMan time is the sum of two consecutive deformation calls, since deploying a single stent across the LPA-daughter junction required two area-manipulation invocations as described in the text.
  \end{tablenotes}
  \end{threeparttable}
\end{table}

\newpage
\subsection{Evaluation of free parameters}
\label{sec:free_params}

In this section, we present the effects of varying key parameters of the model to highlight their roles.

\begin{figure*}[htbp]
  \centering
  \includegraphics[trim=220 190 280 220, clip, width=0.25\textwidth]{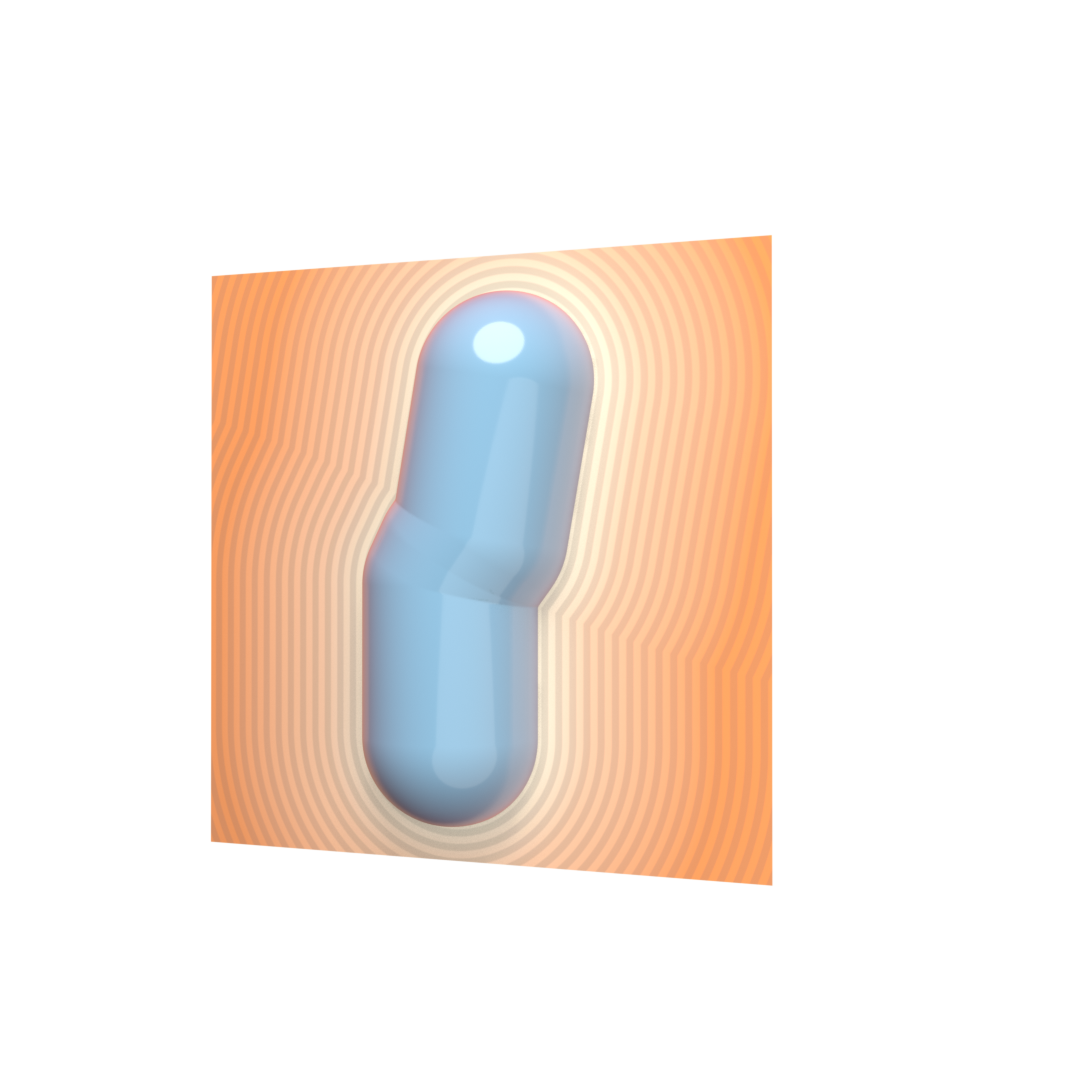}%
  \hspace{0.0\textwidth}%
  \includegraphics[trim=220 190 280 220, clip, width=0.25\textwidth]{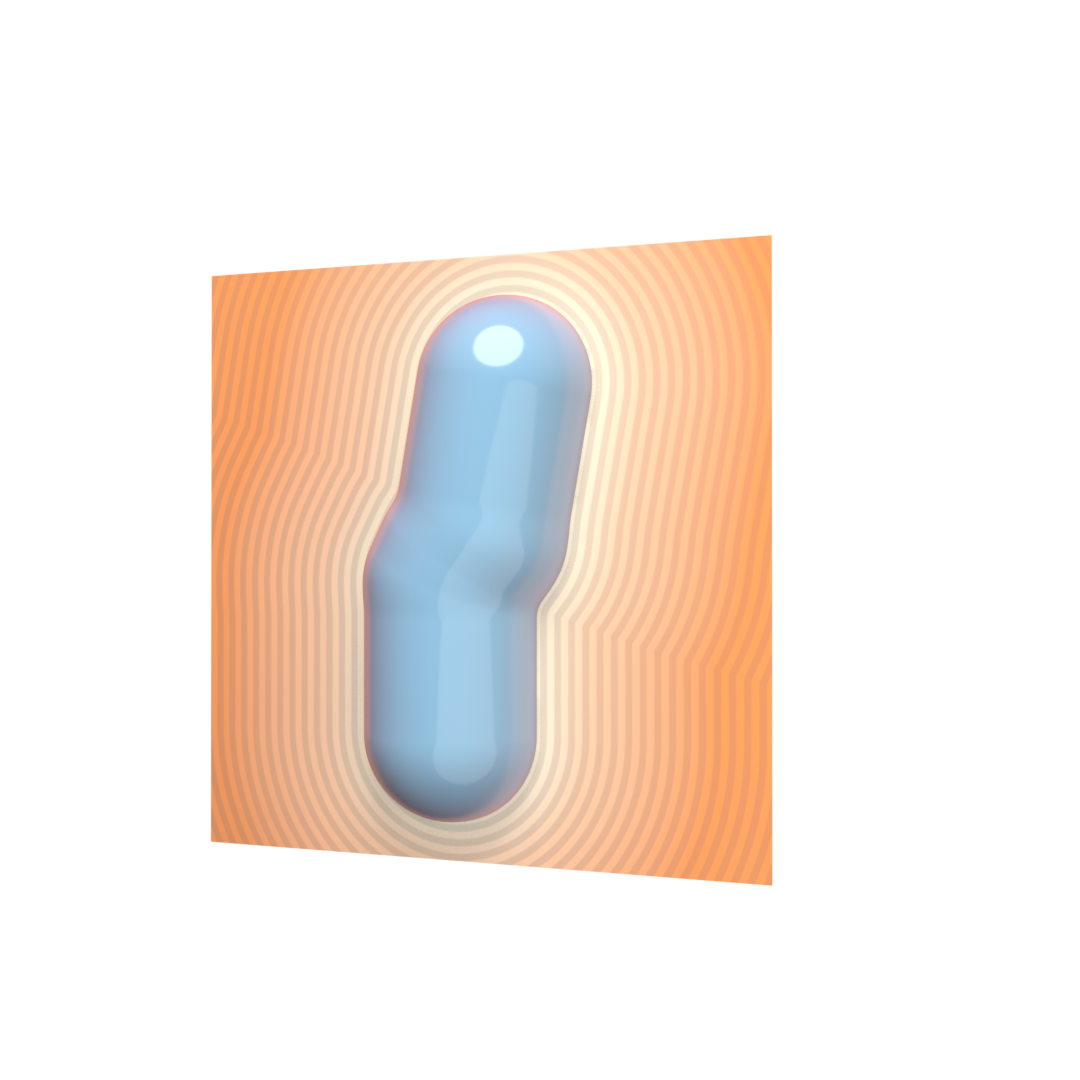}%
  \hspace{0.0\textwidth}%
  \includegraphics[trim=220 190 280 220, clip, width=0.25\textwidth]{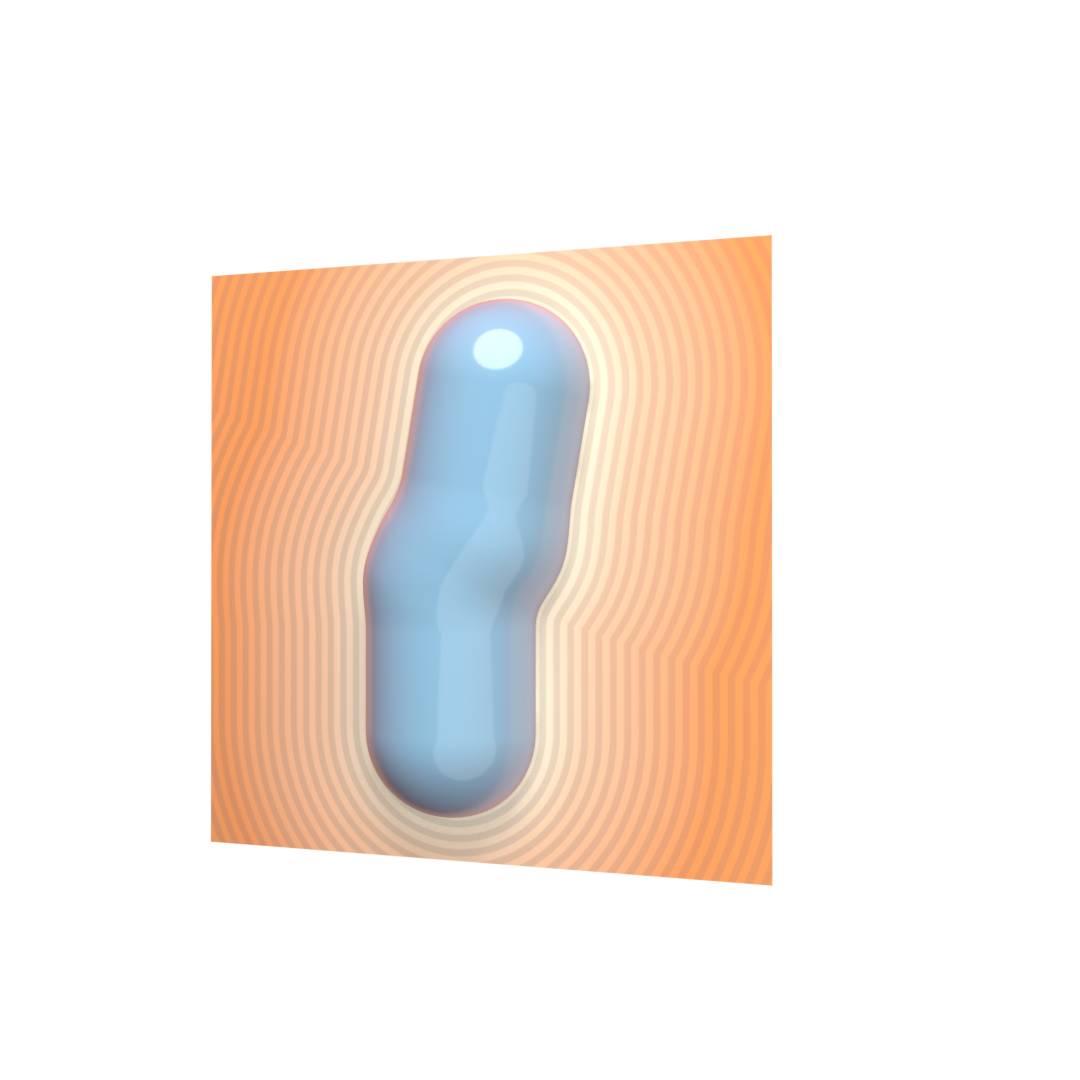}%
  \hspace{0.0\textwidth}%
  \includegraphics[trim=220 190 280 220, clip, width=0.25\textwidth]{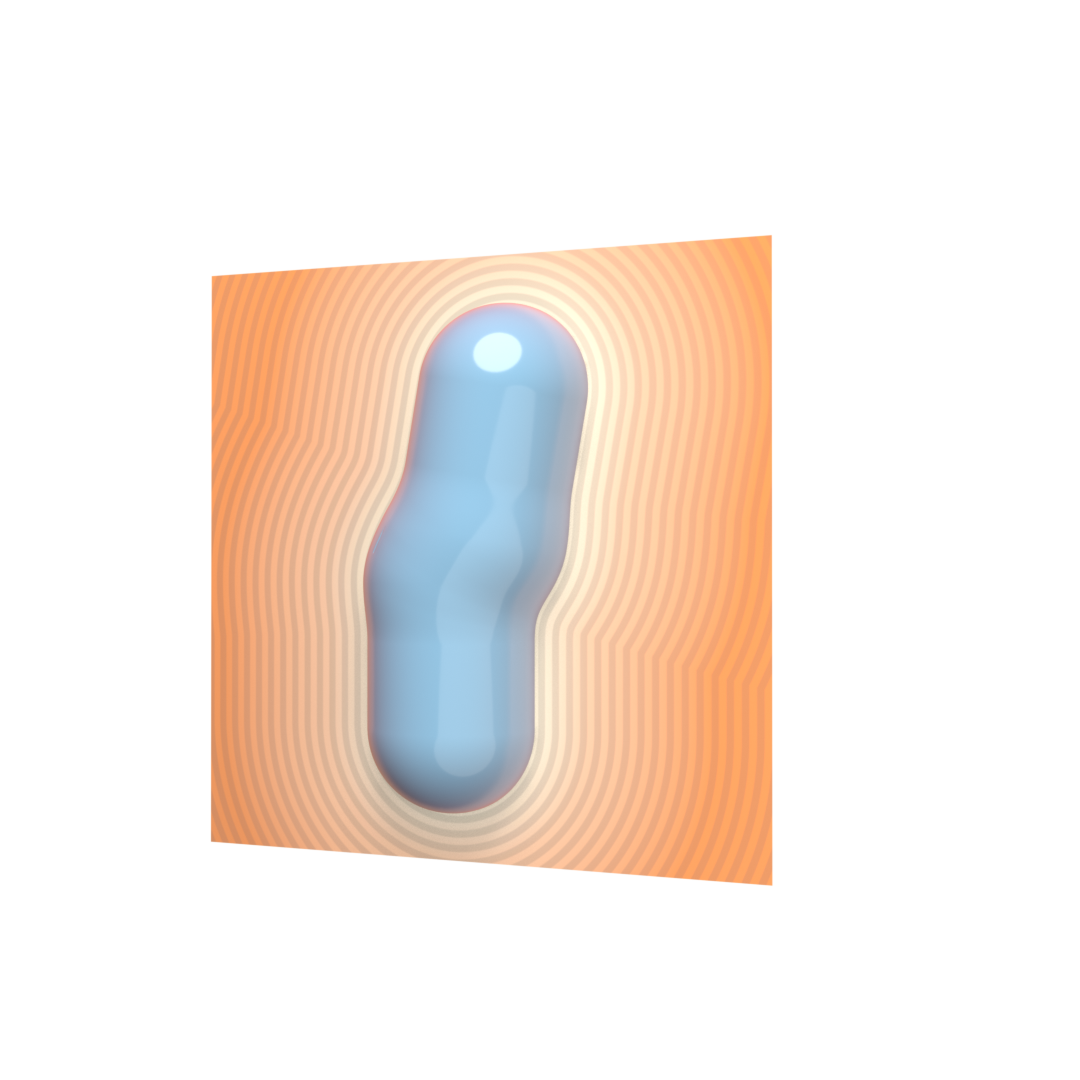}%
  \caption{Visualizations of the SDFs of a synthetic stent shape with a twist in the middle section under varying levels of the smoothing parameter $k$. The synthetic stent model depicted is the union of four near-vertical capsule segments with lengths 2.5mm,0.5mm,0.5mm, and 2.5mm. The in-plane joint angles are $-27$\textdegree, $10$\textdegree{} and $27$\textdegree, which correspond to the maximum ($27$\textdegree) and the 99.7th percentile ($10$\textdegree) of the centerline joint angles observed in our patient-specific vascular models. The effects of the smoothing are further visualized through contours of the SDF in the cross-sectional planes. From left to right: $k = 0.0\textup{mm}, 0.2\textup{mm}, 0.4\textup{mm}, 0.6\textup{mm}$.}
  \label{fig:k_parameter_vary}
\end{figure*}

As mentioned in Section~\ref{sec:smooth_min}, the quadratic smooth minimum function takes in a crucial parameter $k$ that determines the strength of the smoothing. When $k = 0$ one recovers the naive minimum function, while when $k$ is large, it greatly reduces the sharpness of the stent and flattens the creases at the joints. Figure~\ref{fig:k_parameter_vary} shows a secondary consequence common when using smooth minimum functions with increasing smoothing strengths. As we increase $k$, the joint regions between consecutive capsule segments become not only smoother, but also slightly inflated. The effect is much weaker when the angle between two consecutive segments is small, which is almost always the case since we perform a dense resampling of the centerline as explained in Section~\ref{sec:stent_axis}. Therefore, we settled on a value of $k = 0.4$ mm, which strikes a balance between sharpness reduction and feature preservation given the range of centerline joint angles (0\textdegree--27\textdegree) observed in the vascular models we tested.

%% --- Full-width figure: k parameter variation ---

Finally, the distance of influence ($d_{\mathrm{infl}}$) is a free parameter that controls how far the displacement field extends beyond the immediate contact region of the stent. The motivation for nonzero displacements in the surrounding zone is qualitative: in vivo stent deployments visibly displace not only the targeted lumen but also adjacent vascular structures. Just as near-stent vessel wall tissue compresses around the expanding lumen rather than being absorbed by it, the near-stent vessel mesh surface should deform with the expanding lumen rather than be abruptly clipped against it. For stents deployed near junctions, extending the deformation volumetrically around the stent preserves the original junction mesh and avoids erasing the smooth transitions just adjacent to the junction. Our construction is purely geometric, however, as $d_{\mathrm{infl}}$ sets the spatial reach of the fall-off kernel, with no claims about volume preservation, contact mechanics, or the constitutive properties of the vessel wall and surrounding tissue.

\begin{figure*}[htbp]
  \centering
  \includegraphics[width=\textwidth]{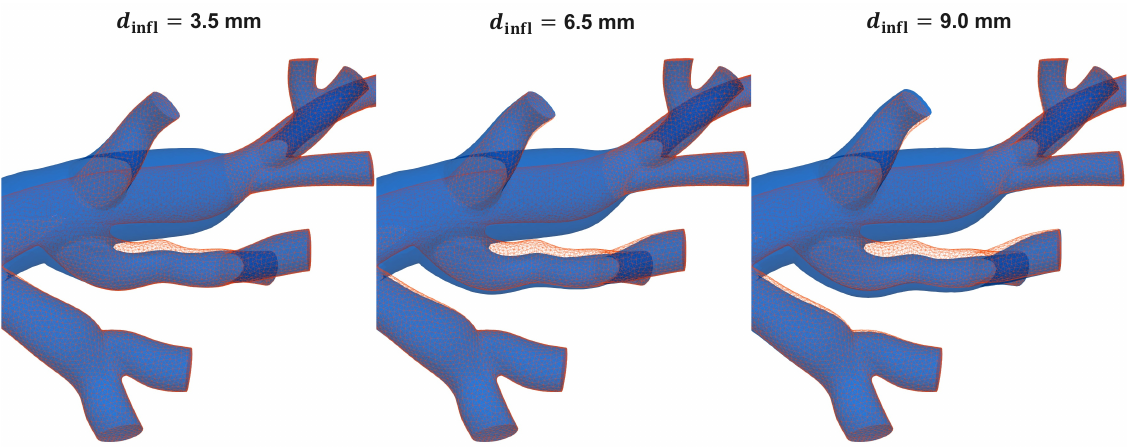}
  \caption{Effect of the distance-of-influence parameter ($d_{\mathrm{infl}}$) on the virtually stented geometry of a representative ToF patient (TOF-1). Each panel overlays the post-stent model (solid blue surface) with the pre-operative geometry (red wireframe) at $d_{\mathrm{infl}} = 3.5\,\mathrm{mm}$ (left), $6.5\,\mathrm{mm}$ (center), and $9.0\,\mathrm{mm}$ (right). At $d_{\mathrm{infl}} = 3.5\,\mathrm{mm}$, the deformation is confined to a narrow shell around the stent, leaving neighboring branches nearly unchanged from their pre-operative positions and constricting the inter-vascular space at the junction. At $d_{\mathrm{infl}} = 9.0\,\mathrm{mm}$, surrounding vessels are displaced outward extensively, distorting anatomy distal to the stent site. At $d_{\mathrm{infl}} = 6.5\,\mathrm{mm}$, junction geometry is preserved while distant branches remain largely undisturbed. Regions where the red wireframe is visible outside the blue surface indicate outward displacement of the vessel wall; regions where the two coincide indicate no deformation.}
  \label{fig:doi-varying-comparisons}
\end{figure*}

\newpage
Figure~\ref{fig:doi-varying-comparisons} illustrates how $d_{\mathrm{infl}}$ directly controls the behavior described above. We selected a value of $d_{\mathrm{infl}} = 6.5$~mm based on the following considerations: if $d_{\mathrm{infl}}$ is too small, deformation is confined too tightly and nearby inter-vascular spaces are unrealistically collapsed against unmoving adjacent surfaces; if $d_{\mathrm{infl}}$ is too large, distant vessels are excessively displaced. The critical length scale that affects this trade-off is the local inter-vascular distance between the stented vessel and adjacent structures. Across both cohorts, junction spacings near typical stenting sites are on the order of several millimeters, and our value of $d_{\mathrm{infl}} = 6.5$~mm produces a deformation reach within this characteristic range.

The selected value was fixed throughout our subsequent experiments rather than re-tuned per patient, since per-patient numerical optimization of $d_{\mathrm{infl}}$ against ground-truth post-operative 3D anatomy was infeasible in our study: post-operative volumetric imaging was often unavailable for CHD patients, and when available for pediatric patients, was usually taken several years after the pre-operative imaging. For our cohort, post-operative models created from long-term follow-up imaging failed to serve as a controlled, static reference, as patient growth introduced significant geometric variations both in anatomic scale and in distal artery geometry. Empirically, a single fixed value of $d_{\mathrm{infl}} = 6.5$~mm generalized well across all six patients in our validation cohort (Section~\ref{sec:patient_validations}) when compared against post-stent angiograms, balancing junction preservation and plausible deformation of distant vessels. Future work coupling our geometric method with patient-specific material characterization and FE tissue models could refine $d_{\mathrm{infl}}$ or replace it with a spatially varying field.

%% --- Full-width figure: DOI comparison ---

\subsection{Patient-specific validation}
\label{sec:patient_validations}

To further validate the effectiveness of SDFStent in real-world clinical scenarios, we collected and de-identified clinical data for a cohort of six patients who each had a moderate to severe stenotic lesion due to CHD and underwent stenting in the cardiac catheterization laboratory ("cath lab") at the Lucile Packard Children's Hospital. For each patient we collected the pre-operative volumetric imaging data in the form of CT or MRI, clinical reports documenting the procedures performed, along with the intra-operative planar angiograms and catheter-based pressure measurements taken immediately before and after the stent deployment. The cohort included three tetralogy of Fallot (ToF) patients (TOF-1, TOF-2, TOF-3) and three coarctation of the aorta (CoA) patients (COA-1, COA-2, COA-3). The ToF patients each had a stenosis in one of the main branch pulmonary arteries, i.e., the right pulmonary artery (RPA) or the LPA. The CoA patients each had a coarctation in the descending aorta. For the ToF patients, we additionally collected the pre-operative and post-operative pulmonary flow split percentages derived from lung-perfusion scans.

All six pre-operative anatomic models were constructed using the standard modeling workflow of the SimVascular simulation suite, consisting of imaging visualization, path identification, cross-section segmentation, surface lofting, junction smoothing and centerline ROM extraction. The three ToF models were constructed from the volumetric imaging data, while the three CoA models and their underlying imaging were collected from a previous study by Nair et al.~\cite{nair_non-invasive_2024} and remeshed with centerlines re-extracted for consistency with the ToF models. Additional attention was paid to ensure that the dimensions of the stenotic regions of the models accurately reflect the measurements recorded in the clinical reports. Lastly, centerline extraction was performed using SimVascular's VMTK-based algorithm for building 1D ROMs~\cite{pfaller_automated_2022, antiga_image-based_2008}.

To perform virtual stenting on our cohort, we extract the diameter and length specifications of the stents directly from the clinical catheterization reports. Balloon-expandable stents undergo axial length reduction upon radial expansion, a phenomenon known as foreshortening, which can reach $\sim$20$\%$ depending on the stent cell geometry and the deployed diameter~\cite{ewert_cp_2005, peters_role_2009}. For the CoA patients, the brand and model of the stent used is the NuMED Covered CP Stent (NuMED Inc., Hopkinton, NY, USA), for which published manufacturer data of foreshortening percentages exists, and we adopt these values here. For the ToF patients, the specific foreshortening percentages were unknown so an average value of 10\% was applied for all three patients. During deployment, the patient's anatomic model and centerline are loaded into the tool, and the position of the distal end of the stent relative to the native vessel was determined by eye with reference to the intra-operative planar angiograms. The virtual stent was then inflated in real time until it reached the clinically recorded final diameter. No algorithmic registration or reference was made to post-operative volumetric imaging.

\begin{figure*}[htbp]
  \centering
  \includegraphics[width=\textwidth]{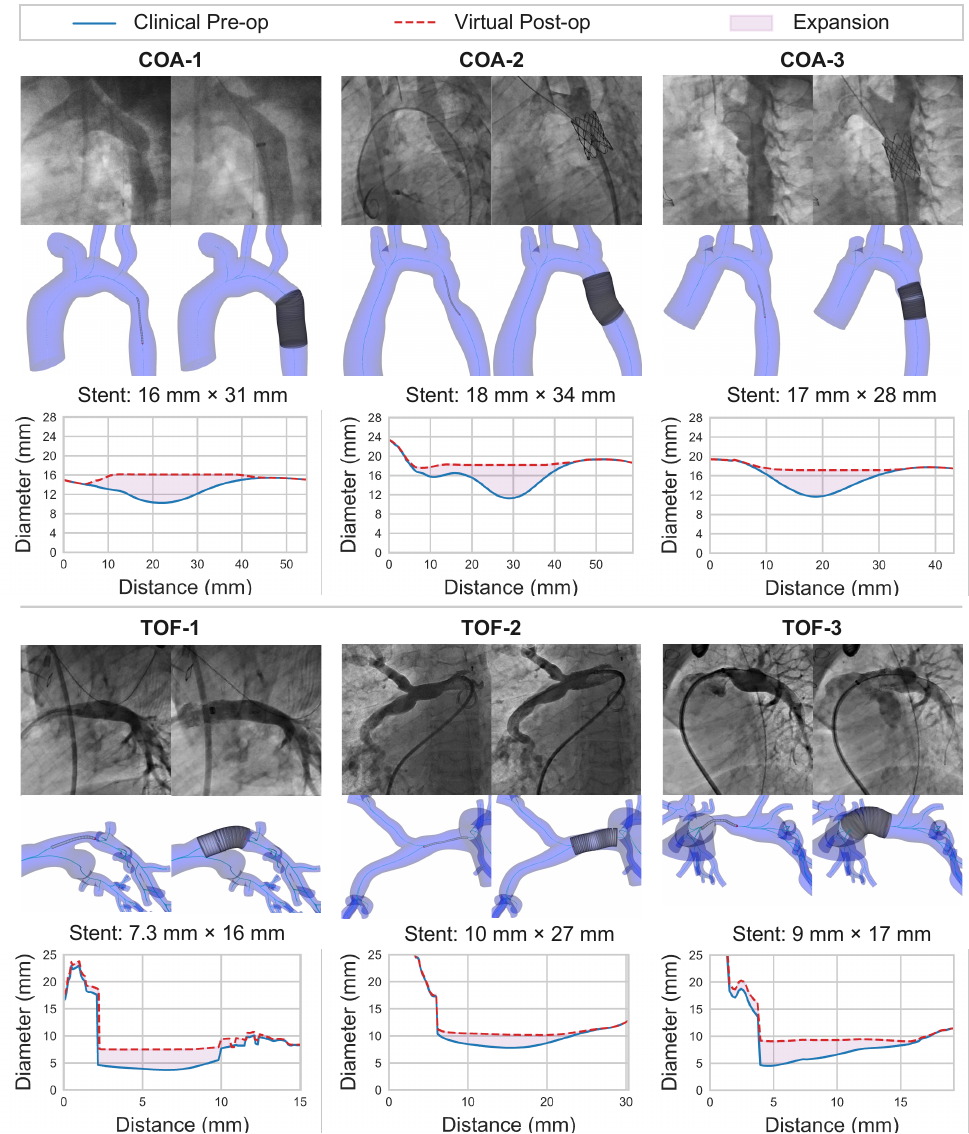}
  \caption{Comparison of clinical fluoroscopy and the corresponding virtual stenting simulations for a cohort of three CoA patients (top) and three ToF patients (bottom). The diameter and length of each deployed virtual stent were directly extracted from the catheterization report. Stent dimensions are reported as diameter $\times$ length, where the length is the nominal value before foreshortening. TOF-1 was reported to have an hourglass stent shape, with a maximum diameter of 8 mm and a minimum of 6.6 mm, so an average of 7.3 mm was used. Side-by-side comparison shows close visual agreement between the in silico and in vivo vessel configurations, achieved in real time without any post-processing.}
  \label{fig:morphological-agreement-virtual-stent-vs-angiogram}
\end{figure*}

Figure~\ref{fig:morphological-agreement-virtual-stent-vs-angiogram} compares the six pairs of pre-stent and virtual post-stent models with reference post-stent planar angiograms. All post-stent model surfaces from our interactive editing sessions remained valid, watertight manifolds and were free of self-intersections, as verified in MeshLab~\cite{cignoni_meshlab_2008} using the ``Compute Topological Measures'' and ``Select Self Intersecting Faces'' filters.

%% --- Full-width figure: morphological agreement ---

Next, we performed CFD simulations using the svMultiPhysics~\footnote{\url{https://github.com/SimVascular/svMultiPhysics} (accessed 13 May 2026)} FE solver in SimVascular. Specifically, \mbox{svMultiPhysics} solves the incompressible Navier--Stokes equations using linear tetrahedral elements with equal-order interpolation for velocity and pressure, stabilized via the residual-based variational multiscale (RBVMS) method~\cite{bazilevs_variational_2007}. Time integration uses the generalized-$\alpha$ method~\cite{jansen_generalized-alpha_2000} with spectral radius $\rho_\infty = 0.2$, and the resulting nonlinear system is solved at each timestep via Newton--Raphson iteration. To reduce computational cost and avoid introducing additional layers of physics that would confound the geometric comparison, we performed rigid-wall, steady-state CFD simulations for all six pairs of patient-specific models. We tetrahedralized each pair of pre-stent and post-stent surfaces into volume meshes directly, without further processing, using the open-source meshing software TetGen~\cite{si_tetgen_2015}. The element sizes were fixed within each surface pair and chosen to produce 1.5 to 2.5 million elements, which is sufficient for our steady-state simulations. Table~\ref{tab:mesh-specs} summarizes the final mesh specifications for the cohort.

For five of the patients, the volumetric imaging data were in the form of 4D-Flow MRI, therefore inflow waveforms were extracted using Arterys (Arterys Inc., San Francisco, CA, USA), which implements 4D flow analysis methods described in~\cite{vasanawala_congenital_2015}. For TOF-1, inflow was unavailable, so we uniformly scaled the waveform from TOF-3 according to the ratio of their cardiac output, noting that TOF-1 and TOF-3 were both pediatric patients ($\leq$ 4 years) with severe pulmonary regurgitation. Since pressure drop values in the catheterization reports were measured clinically as differences in peak systolic pressures, we extracted the maximum value of each waveform and broadcast it as the steady inflow BC.

\begin{table*}[htbp]
  \centering
  \caption{Volume mesh specifications for the study cohort.}
  \label{tab:mesh-specs}
  \setlength{\tabcolsep}{8pt}
  \begin{tabular}{l r r r r r}
    \toprule
    & & \multicolumn{2}{c}{\textbf{Pre-operative}} & \multicolumn{2}{c}{\textbf{Post-operative}} \\
    \cmidrule(lr){3-4}\cmidrule(lr){5-6}
    \textbf{Patient ID} & \textbf{Element size (cm)} & \textbf{\# Nodes} & \textbf{\# Elements} & \textbf{\# Nodes} & \textbf{\# Elements} \\
    \midrule
    COA-1 & 0.08 & 263,742 & 1,559,980 & 276,712 & 1,641,676 \\
    COA-2 & 0.08 & 361,278 & 2,169,989 & 377,633 & 2,269,290 \\
    COA-3 & 0.08 & 438,829 & 2,648,964 & 449,625 & 2,716,821 \\
    TOF-1 & 0.04 & 283,721 & 1,634,177 & 299,427 & 1,735,847 \\
    TOF-2 & 0.06 & 371,623 & 2,190,458 & 376,797 & 2,224,432 \\
    TOF-3 & 0.06 & 232,233 & 1,333,140 & 238,375 & 1,372,924 \\
    \bottomrule
  \end{tabular}
\end{table*}

Our simulation studies were carried out in two phases. In phase one, we prescribed a steady inflow BC with a parabolic velocity profile at each pre-operative model's inlet. We then iteratively tuned the outlet resistance BCs until key hemodynamic tuning targets matched clinical catheterization data. In phase two, we performed simulations on the corresponding post-operative models using identical BCs and simulation parameters, isolating the geometry change from the virtual stent intervention as the only variable.

To select the clinical tuning targets in phase one, we identified hemodynamic metrics essential to the diagnosis and treatment of the two CHDs in our study that were both available in our clinical data and measurable within our anatomic models. For the CoA cohort, we tuned outlet resistances until the simulated ascending aorta (AAo) and descending aorta (DAo) pressures matched the clinical measurements. For the ToF cohort, we tuned outlet resistances until the simulated main pulmonary artery (MPA) pressure and RPA/LPA flow split percentages matched the measurements. All tuning targets were matched to the nearest integer, corresponding to the lowest common precision of the catheterization reports.

In addition to the tuning targets, we recorded the pressure drop across the stenosis $\Delta P = P_\mathrm{prox} - P_\mathrm{dist}$ as a verification metric that was not constrained during tuning. The trans-stenotic pressure gradient is a primary clinical indicator for intervention in both CoA and ToF, and reporting it under one unified definition across cohorts allows the same metric to validate our predictions for both diseases.

Each named pressure was extracted from the simulation at a standardized centerline cross-section that mirrors a clinical catheter position. AAo for CoA and MPA for ToF are placed at the midpoint, by centerline arclength, between the inlet and the first vascular junction. DAo for CoA is placed two vertebral spaces above the diaphragm, which is the same anatomic landmark used by Nair et al.~\cite{nair_non-invasive_2024} for these CoA patients. The pressure-drop endpoints are cohort-specific. For the CoA cohort, $P_\mathrm{prox}$ is placed immediately distal to the left subclavian artery junction, and $P_\mathrm{dist}$ coincides with the DAo tuning site. For the ToF cohort, $P_\mathrm{prox}$ coincides with the MPA tuning site, and $P_\mathrm{dist}$ is placed distal to the stenosis and immediately proximal to the next downstream junction. Both $P_\mathrm{prox}$ for CoA and $P_\mathrm{dist}$ for ToF were chosen in consultation with clinicians to be distal enough from the stenosis center to clear the edge of the stent where the vessel returns to its native diameter, while proximal enough to avoid the flow irregularities introduced by nearby vascular junctions.

\begin{figure*}[htbp]
  \centering
  \includegraphics[width=\textwidth]{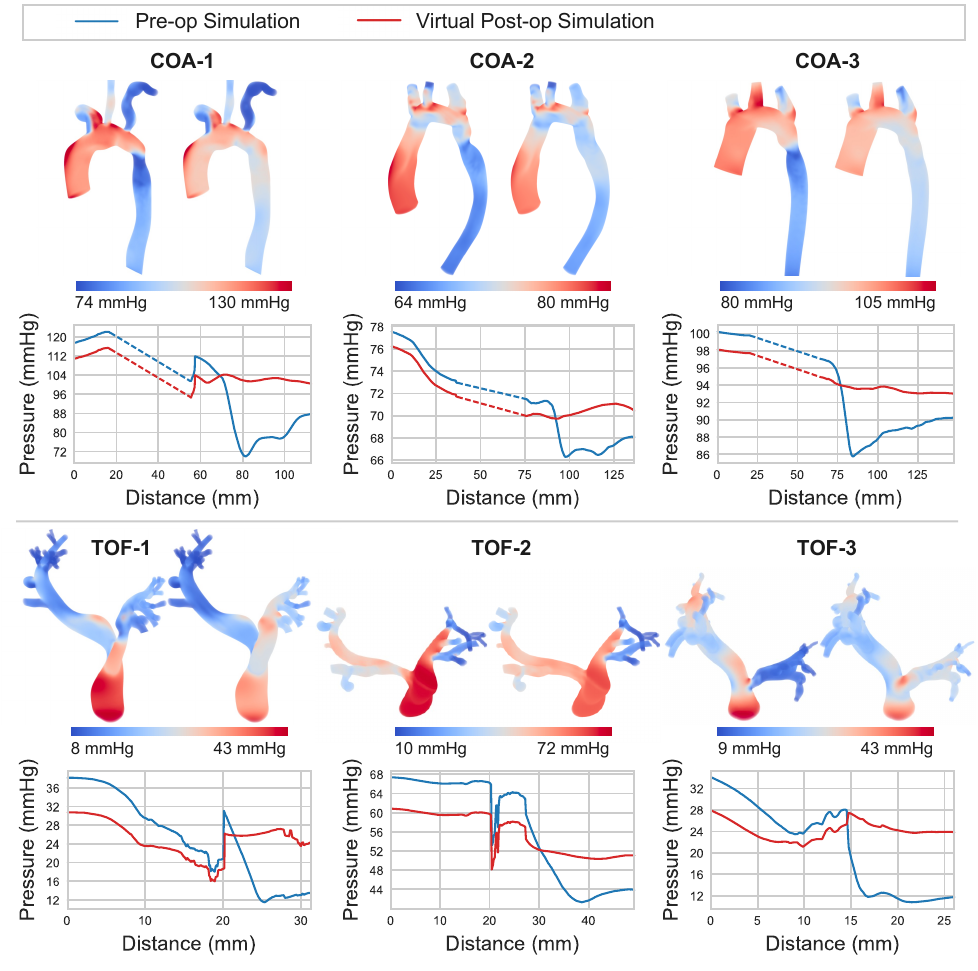}
  \caption{Pre-operative and virtual post-operative pressure distributions from CFD simulations
for the CoA (top) and ToF (bottom) cohorts.
For each patient, 3D pressure volumetric renderings are shown alongside centerline pressure profiles.
For the CoA cases, cross-section-averaged pressures are extracted from the AAo measurement
site to the beginning of the aortic arch and from the proximal pressure drop measurement site at the left subclavian artery (LSCA) junction to the DAo measurement site; the intervening arch segment is omitted due to the irregular cross-sections introduced by the branching arch vessels, with dashed lines indicating linear interpolation across this gap.}
  \label{fig:cfd-pressure-comparisons}
\end{figure*}

All simulations were advanced for 1000 timesteps with a step size of $5 \times 10^{-4}$~s, yielding CFL numbers of approximately 1.5 and a total physical time of 0.5~s, corresponding to 3--4 flow-throughs in each simulation domain. The linear systems arising at each Newton--Raphson iteration were solved using the bi-partitioned (BIPN) method~\cite{esmaily-moghadam_bi-partitioned_2015}, which splits the velocity-pressure system into momentum and pressure blocks solved with GMRES and CG respectively, paired with a diagonal resistance-based preconditioner to handle the ill-conditioning introduced by the resistance outlet BCs. Residual convergence of the steady-state solution was verified in all cases. Each simulation took approximately 1 hour of wall-clock time on two 48-core compute nodes of the Stampede3 cluster. Figure~\ref{fig:cfd-pressure-comparisons} shows our simulated peak systolic pressure distribution results.

Pressures were temporally averaged over the last 200 timesteps of simulation, then spatially averaged across the cross-section orthogonal to the vessel centerline at each centerline point. Spatial standard deviations over the cross-sections typically remained below 4~mmHg across all post-operative measurement sites, and typically higher near junctions, confirming that cross-sectional averages adequately capture the local hemodynamic pressures in closed tubular measurement sites such as AAo, DAo, and MPA, and are therefore comparable to clinical single-point catheter pressure measurements despite variations of the catheter tip position within the cross-section during the procedure.

Across all six patients, we observe significant relief in pathological pressure drops across the focal stenoses after virtual stenting, as well as mild reductions of absolute upstream pressures. For ToF patients, we also observe slight reductions in pressures in the non-stenosed pulmonary branches, which suggests a shift in flow distribution towards the stented branch and overall reduction in total pulmonary vascular resistance (PVR).

Tables~\ref{tab:pressure-results-coa} and~\ref{tab:pressure-results-tof} present a quantitative comparison of clinical and simulated hemodynamic outcomes for the CoA and ToF cohorts, respectively. We focus our results discussion on pressure drop across the stenosis, which is clinically regarded as much more reliable than absolute pressure measurements for validation purposes: it is intrinsically a difference between two absolute pressures and thus partially cancels out pressure drifts arising from changing patient conditions, such as anesthesia depth and cardiac output variations throughout the procedure. This is especially relevant for the CoA cohort, where the reported femoral artery standing wave phenomenon~\cite{nichols_mcdonalds_2011} and changes in anesthesia depth between pre- and post-stent measurements shifted the absolute AAo and DAo readings in our clinical data, motivating our choice to report CoA post-operative results in terms of pressure drop only.

Across all six patients, simulated post-operative pressure drops agreed with clinical measurements within 4~mmHg with a mean error of 2~mmHg. For the CoA cohort, the simulated post-operative pressure drops were 4, 0, and 2~mmHg, against clinical values of 1, 0, and 0~mmHg. For the ToF cohort, the simulated post-operative pressure drops were 4, 10, and 4~mmHg, against clinical values of 0, 7, and 6~mmHg. Both cohorts show post-stent gradient relief that closely tracks the clinically observed drop, providing direct evidence that the SDFStent geometry alterations propagate consistently into downstream CFD predictions of stenosis relief.

Pre-operative pressure drops were also reported as a verification metric not constrained during BC tuning. Simulated pre-operative values underestimated clinical measurements by 3--7~mmHg in CoA and overestimated by 2--7~mmHg in ToF. The largest relative gap was COA-2 (simulated 3~mmHg versus clinical 10~mmHg), which we discuss in Section~\ref{sec:discussion} in connection with measurement-site choices in cath lab procedures. Pre-operative AAo, DAo, and MPA pressures match clinical values exactly by construction via BC tuning; post-operative MPA pressures for the ToF cohort agreed with clinical measurements within 6~mmHg (mean error 4~mmHg). Flow split predictions were generally consistent with clinical measurements for TOF-1 and TOF-2, though a larger discrepancy was observed for TOF-3 (simulated RPA/LPA of 78/22 versus clinical 63/37), which may reflect the flow-mediated adaptation of distal PVR that a fixed BC model does not capture.

In clinical practice, a resting peak-to-peak aortic pressure gradient exceeding 10~mmHg, together with anatomic and functional findings, is a crucial threshold when considering catheter intervention for CoA patients~\cite{stout_2018_2019}. Similarly, a pressure gradient exceeding 20~mmHg across a branch pulmonary artery stenosis, particularly in the presence of flow imbalance, is a common hemodynamic criterion for catheter intervention in ToF patients~\cite{prakoso_main_2026}. The simulated post-operative gradients reported above all fell well below their respective intervention thresholds, in agreement with the corresponding clinical post-stent measurements. Simulation therefore correctly predicted intervention success across all six patients, in the sense that post-stent gradients were no longer indicative of further intervention.

%% --- Full-width figure: pressure comparison ---

\begin{table*}[htbp]
  \centering
  \begin{threeparttable}
  \caption{Clinical versus simulated hemodynamic metrics for the CoA cohort.}
  \label{tab:pressure-results-coa}
  \setlength{\tabcolsep}{4pt}
  \small
  \begin{tabular}{l r r r r r r r r}
    \toprule
    & \textbf{AAo (mmHg)}
    & \textbf{DAo (mmHg)}
    & \multicolumn{6}{c}{\textbf{Pressure Drop (mmHg)}} \\
    \cmidrule(lr){2-2}\cmidrule(lr){3-3}\cmidrule(lr){4-9}
    & \textbf{Pre-op}
    & \textbf{Pre-op}
    & \multicolumn{3}{c}{\textbf{Pre-op}}
    & \multicolumn{3}{c}{\textbf{Post-op}} \\
    \cmidrule(lr){4-6}\cmidrule(lr){7-9}
    \textbf{Patient ID}
      & \textbf{Clin.=Sim.}
      & \textbf{Clin.=Sim.}
      & \textbf{Clin.} & \textbf{Sim.} & \textbf{Diff}
      & \textbf{Clin.} & \textbf{Sim.} & \textbf{Diff} \\
    \midrule
    COA-1 & 118 & 88 & 30 & 24 & $-6$ & 1 & 4 & $+3$ \\
    COA-2 & 78  & 68 & 10 & 3  & $-7$ & 0 & 0 & 0    \\
    COA-3 & 100 & 90 & 10 & 7  & $-3$ & 0 & 2 & $+2$ \\
    \bottomrule
  \end{tabular}
  \begin{tablenotes}[flushleft]
    \footnotesize
    \item Pre-operative clinical and simulated AAo and DAo pressures are identical after BC tuning (Clin.\ = Sim.); simulated pressure drops, measured as the difference between $P_\mathrm{prox}$ and $P_\mathrm{dist} \equiv$ DAo, are a verification metric and not constrained during BC tuning. Post-operative differences (Diff) are defined as Simulation $-$ Clinical.
  \end{tablenotes}
  \end{threeparttable}
\end{table*}

\begin{table*}[htbp]
  \centering
  \begin{threeparttable}
  \caption{Clinical versus simulated hemodynamic metrics for the ToF cohort.}
  \label{tab:pressure-results-tof}
  \setlength{\tabcolsep}{4pt}
  \small
  \begin{tabular}{l r r r r r r r r r r r r r r}
    \toprule
    & \multicolumn{4}{c}{\textbf{MPA Pressure (mmHg)}}
    & \multicolumn{4}{c}{\textbf{Flow Split (RPA/LPA \%)}}
    & \multicolumn{6}{c}{\textbf{Pressure Drop (mmHg)}} \\
    \cmidrule(lr){2-5}\cmidrule(lr){6-9}\cmidrule(lr){10-15}
    & \textbf{Pre-op}
    & \multicolumn{3}{c}{\textbf{Post-op}}
    & \textbf{Pre-op}
    & \multicolumn{3}{c}{\textbf{Post-op}}
    & \multicolumn{3}{c}{\textbf{Pre-op}}
    & \multicolumn{3}{c}{\textbf{Post-op}} \\
    \cmidrule(lr){3-5}\cmidrule(lr){7-9}\cmidrule(lr){10-12}\cmidrule(lr){13-15}
    \textbf{Patient ID}
      & \textbf{Clin.=Sim.}
      & \textbf{Clin.} & \textbf{Sim.} & \textbf{Diff}
      & \textbf{Clin.=Sim.}
      & \textbf{Clin.} & \textbf{Sim.} & \textbf{Diff}
      & \textbf{Clin.} & \textbf{Sim.} & \textbf{Diff}
      & \textbf{Clin.} & \textbf{Sim.} & \textbf{Diff} \\
    \midrule
    TOF-1 & 38 & 25 & 31 & +6   & 71/29 & 56/44 & 61/39 & $-5$    & 18 & 25 & $+7$  & 0 & 4  & $+4$  \\
    TOF-2 & 67 & 64 & 61 & $-3$ & 45/55 & 47/53 & 47/53 & 0     & 21 & 23 & $+2$  & 7 & 10 & $+3$  \\
    TOF-3 & 34 & 30 & 28 & $-2$ & 87/13 & 63/37 & 78/22 & $-15$   & 20 & 22 & +2  & 6 & 4  & $-2$  \\
    \bottomrule
  \end{tabular}
  \begin{tablenotes}[flushleft]
    \footnotesize
    \item Pre-operative clinical and simulated MPA pressures and flow splits are identical after BC tuning (Clin.\ = Sim.); simulated pressure drops, measured as the difference between $P_\mathrm{prox} \equiv$ MPA and $P_\mathrm{dist}$, are a verification metric and not constrained during BC tuning. Post-operative differences (Diff) are defined as Simulation $-$ Clinical.
  \end{tablenotes}
  \end{threeparttable}
\end{table*}

%% =====================================================================
%%  4.  DISCUSSION
%% =====================================================================
\section{Discussion} 
\label{sec:discussion}

SDFStent addresses a long-standing bottleneck for morphology-driven hemodynamic studies of stent interventions. It delivers precise parametric control over stent-induced shape changes while remaining time-efficient and simple to operate, whereas previous methods compromise on at least one of these three criteria.

Another strength of SDFStent is its modular nature. The 1D fall-off profile function can be independently substituted with other monotone decreasing functions to emulate different material characteristics. The two adjustable free parameters in the model (smoothing factor $k$ and distance of influence $d_{\mathrm{infl}}$) directly control the deformation process with intuitive effects, and they are decoupled from the diameter and length of the virtual stent. This is in contrast to methods like regularized Kelvinlets, where the regularization kernel couples smoothness and deformation strength, making precise shape adjustments less straightforward.

While interactive sculpting tools like MeshMixer~(Autodesk Inc., San Rafael, CA, USA) are often not easily scriptable, and compute-heavy methods like XPBD are not easily made interactive, SDFStent offers the rare advantage of being both interactive and scriptable. A non-expert user could first prototype by hand using the GUI with relative ease. Then, once the stent locations are confirmed through trial and error, the GUI-less API can be deployed at scale without user interaction to parametrically generate combinations, degrees, and extents of stent inflation. This holds significant value for data-hungry downstream applications like machine learning methods. As data-driven ROMs such as neural surrogate solvers~\cite{du_deep_2022} and amortized inference frameworks~\cite{choi_falconbc_2026} continue to mature, coupling them with our real-time geometry editing method could bring interactive cardiovascular digital twins~\cite{corral-acero_digital_2020} closer to clinical use: a clinician could adjust a virtual stent and immediately observe the predicted hemodynamic response within a single planning session.

From a practical standpoint, SDFStent offers ease of use for simulation engineers and clinicians alike, requiring no specialized computing hardware such as VR headsets or GPUs, and it runs in real time on average consumer machines. One can readily and reliably perform virtual stenting on any vascular surface meshes in the VTP format; for instance, it can readily work on all 316 models from the Vascular Model Repository~\footnote{\url{https://www.vascularmodel.com} (accessed 13 May 2026)}. Moreover, for surface meshes that do not have an accompanying centerline, centerlines can be readily extracted via automated tools available in SimVascular~\cite{pfaller_automated_2022} or VMTK. Additionally, one could manually specify the axis of the virtual stent by constructing a 3D centerline path using any existing geometric curve creation tools, so long as the final input to the method is a simple connected path. In fact, we have already extended the SDFStent 3D Slicer module to allow the use of arbitrary path lines as stent axes, which makes it possible to relax its curvature or perturb its relative position within the lumen. This extension is particularly relevant for closed-cell stents, which can exhibit up to twice the amount of straightening of open-cell designs after inflation~\cite{liao_three-dimensional_2004}.

\subsection{Clinical implications}
\label{sec:discussion-clinical}

The broader motivation for our work is that the hemodynamic consequences of a stent intervention are often multi-factored and non-intuitive. Relieving a stenosis can shift flow distributions across the vascular tree, unmask gradients that were not previously dominant, or alter downstream resistance through flow-mediated vasodilation. Our ToF cohort illustrates this well: the three patients had comparable pre-operative pressure drops (18, 21, and 20~mmHg simulated), yet their post-operative flow-distribution outcomes differed considerably. TOF-1 and TOF-3 saw substantial flow redistribution, while TOF-2's split remained essentially unchanged. CFD simulation is one practical way to anticipate these coupled effects, and capturing the post-operative geometry change, though not sufficient to achieve full predictive accuracy, is a necessary step. By making the geometry step take seconds rather than hours of manual editing, SDFStent lowers the practical cost of virtually testing intervention scenarios, so that decisions about whether, where, and how to intervene can be informed by predicted hemodynamics beforehand in addition to intra-procedural judgment.

Our validation experiments using real clinical data from both cohorts support the practical deployment of the method for clinical studies: using only text descriptions of the procedures and 2D angiograms as reference, the resulting in silico models showed close visual agreement with the corresponding intra-operative angiograms (Figure~\ref{fig:morphological-agreement-virtual-stent-vs-angiogram}), and the downstream CFD simulations produced post-operative pressure drops that closely matched clinical measurements. In fact, SDFStent has already been applied to ongoing intervention planning for Fontan conduit upsizing cases at the Lucile Packard Children's Hospital, demonstrating its utility as a clinical tool. Even in the absence of more sophisticated BCs such as RCR~\cite{vignon-clementel_outflow_2006} or BC adaptation techniques~\cite{yang_adaptive_2016, lan_virtual_2022}, the simulations correctly predicted post-stent gradients to be well below intervention thresholds for both aortic coarctation and branch pulmonary artery stenosis. These results suggest that for binary clinical decision-making, i.e., determining whether a patient's hemodynamic parameters fall above or below an actionable cutoff, a geometrically sufficiently accurate post-stent model paired with simplified BCs can preserve the correct clinical classification. In our validation workflow, iterative BC tuning currently accounts for the majority of the total workflow time since each iteration incurs a full simulation run, with a total of 5--6 iterations needed. Subsequent studies that integrate fast BC tuning methods such as~\cite{choi_falconbc_2026} could enable a single-day turnaround for the full rigid-wall, steady-state virtual intervention workflow, providing a rapid preview that informs active clinical planning.

\subsection{Limitations and future work}
\label{sec:discussion-limitations}

One limitation of SDFStent is that it imposes prescribed, empirically tuned fall-off profiles to achieve the desired deformation in both the edited lumen and neighboring lumens. Although this is a step forward from methods that ignore nearby mesh surfaces, it does not solve the underlying solid-mechanics and two-way contact problem between stent and vessel wall. As a result, the predicted displacements are not guaranteed to match real material constitutive behaviors. To bridge this gap, future work could try fitting parametric forms of fall-off profiles using FE calibration simulations of stent contact with thin-walled hyperelastic tubes embedded in ambient tissue. However, the scarcity of pre-operative and post-operative 3D image data pairs and the high variability of patient-specific and location-specific material properties would remain the biggest challenge.

Our CFD validation used rigid-wall, steady-state simulations, which do not capture vessel compliance or pulsatile flow dynamics. Steady inflow at peak systole may overestimate instantaneous pressure drops relative to time-averaged pulsatile values, and the absence of wall motion omits compliance-dependent effects such as pulse pressure amplification in the aorta~\cite{nichols_mcdonalds_2011} and transient systolic dilation of the pulmonary artery stenosis, both of which may reduce the instantaneous pressure gradient relative to a rigid-wall prediction. We adopted these simplifications to isolate the geometric effect of the virtual stent from other modeling variables, but incorporating fluid-structure interaction (FSI) and unsteady inflows would provide a more complete hemodynamic assessment of the post-stent anatomy.

A second limitation concerns the consistency of clinical pressure measurements. This is specific to the CoA cohort: the long aortic geometry separates the AAo and the focal stenosis substantially, while in ToF the upstream pressure measurement site (MPA) coincides with the proximal pressure-drop measurement site by design. Catheter-based pressure measurement in the cath lab is not standardized across pre- and post-stent stages, and in our CoA cohort the pre- and post-operative clinical pressure drops were likely measured at different anatomic locations. The clinical pre-operative pressure drops are consistent with a catheter pullback from the AAo to the DAo, which captures the focal coarctation gradient together with diffuse pressure loss along the aortic arch. The clinical post-operative values, at or near zero, are more consistent with a local measurement across the stented region, since an AAo-to-DAo pullback cannot yield such low values under normal arch hemodynamics. In our simulation setup, we adopted a single consistent measurement convention ($P_\mathrm{prox} - P_\mathrm{dist}$) across pre- and post-op for reproducibility, matching the local trans-stenotic measurement. As a result, the simulated CoA pre-operative pressure drops underestimate the clinical values. A fully consistent comparison would require fixed pressure measurement locations clinically, which were not retrospectively available. Future studies pairing virtual stenting with catheter measurements at fixed, pre-specified locations could close this gap.

Another simplification in our CFD setup is that pre-operative outlet BCs are identically applied to the post-operative geometry so that the geometry change is the only variable. Consequently, the residual disagreement between simulated and clinical post-operative metrics in Section~\ref{sec:patient_validations} reflects a composite of geometric error from the virtual stenting method and BC error from the assumption that distal vascular resistance is unchanged. We expect, however, that adopting anatomy-specific post-operative BC adaptations given the expanded geometry, as demonstrated in prior works on pulmonary arteries~\cite{yang_adaptive_2016, lan_virtual_2022}, would reduce rather than increase the residual error. The flow-split discrepancy in TOF-3 is consistent with this picture: relief of a stenotic pulmonary branch typically increases its flow, lowering its PVR through flow-mediated vasodilation and recruitment of the distal vasculature, which in turn shifts flow further toward the stented branch. A frozen pre-operative resistance cannot represent this positive feedback. The agreement reported under the frozen-BC setup is therefore best read as a conservative bound on the geometric component of error, with adaptive-BC integration offering a promising direction for even more accurate predictions.

The real-time behavior reported in our study was measured on surface meshes of approximately 100,000--400,000 triangles (corresponding to volumetric CFD meshes of $\sim$1.5--2.5 million tetrahedra), using vectorized NumPy/SciPy kernels to accelerate proximity queries and linear algebra, and perform coarse-grain culling with an axis-aligned bounding box (AABB) on the stent. This mesh resolution matches most medium-fidelity hemodynamic studies in clinically focused research. However, some studies employ very high-fidelity geometries ($\gg$10 million elements)~\cite{manchester_evaluation_2022}, for instance, to accurately resolve wall shear stress. At those scales, the cost of tree-based nearest-neighbor queries grows superlinearly (e.g., $\approx$O(N log N)), so edit latency can reduce down to single-digit frames per second, reducing user-friendliness. As future work, we will adopt spatial culling methods that better suit the tubular shapes of vascular structures such as Oriented Bounding Boxes (OBBs)~\cite{gottschalk_obbtree_1996} to maintain even better computation time scaling on extremely fine meshes. Additionally, GPU acceleration via Python-to-CUDA frameworks such as NVIDIA Warp~\cite{macklin_warp_2022} could deliver order-of-magnitude speedups with minimal departure from our current Python implementation, at the cost of requiring a CUDA-capable GPU. Alternatively, a preview-and-refine architecture could render a simplified approximation immediately while the full-quality result is computed in the background.

Lastly, since the main focus of our study is the geometric deformation technique, and high-quality paired pre-operative and post-operative patient data are rare, our validation cohort of six patients limits generalizability across the full spectrum of stenosis severities, anatomic variants, and stent designs encountered in clinical practice. In addition, the input pre-operative surfaces were manually reconstructed, so choices about distal inclusion, lumen origin, and branch truncation introduce subjective judgments and variability. As future work, we plan to expand the cohort through continued clinical collaboration, develop standardized segmentation guidelines that ensure consistent topology across pre-operative models, and apply automated segmentation pipelines such as those available in Mimics~(Materialise NV, Leuven, Belgium) or 3D Slicer to reduce operator-dependent variability and better isolate method-specific biases.

%% =====================================================================
%%  5.  CONCLUSION
%% =====================================================================
\section{Conclusion} 
\label{sec:conclusion}  

SDFStent provides a fast, accurate, and accessible solution to a persistent challenge in virtual stent planning: producing simulation-ready post-stent vascular geometries from pre-operative anatomies. Across six retrospective CHD cases, pressure drops predicted from the virtual post-stent models matched clinical catheterization measurements closely with a mean error of 2~mmHg, falling well below intervention thresholds for both aortic coarctation and branch pulmonary artery stenosis. Beyond these cases, SDFStent's interactive speed and scriptable API open a path toward routine inclusion of patient-specific virtual stenting in both clinical planning workflows and large-scale synthetic data pipelines for data-driven hemodynamic modeling.

%% =====================================================================
%%  DATA AVAILABILITY (common in CMPB)
%% =====================================================================
\section*{Data availability}

SDFStent is openly available under a permissive license at \url{https://github.com/SimVascular/svMorph} (source code), \url{https://pypi.org/project/svmorph/} (PyPI package \texttt{svmorph} v0.1.2), and \url{https://github.com/SimVascular/SlicerSimVascular} (3D Slicer module). The six patient-specific anatomic models and their associated de-identified imaging and clinical catheterization data will be deposited in the Vascular Model Repository (\url{https://www.vascularmodel.com}) upon acceptance.

%% =====================================================================
%%  ETHICS STATEMENT (required by CMPB / Elsevier)
%% =====================================================================
\section*{Ethics statement}

This retrospective study was approved by the Stanford University
Institutional Review Board (IRB Protocol \#39377).
All clinical imaging, catheterization, and procedural data were
collected as part of routine clinical care at the Lucile Packard Children's
Hospital and were fully de-identified prior to analysis. Informed
consent was waived by the IRB given the retrospective
nature of the study and the use of fully de-identified data. The
privacy rights of all human subjects were observed throughout the
study, and no patient-identifiable information appears in this
manuscript or its supplementary materials.

%% =====================================================================
%%  CRediT AUTHORSHIP CONTRIBUTION STATEMENT (required by CMPB)
%% =====================================================================
\section*{CRediT authorship contribution statement}

\textbf{Bohan J. Li}: Conceptualization, Methodology, Software, Investigation, Validation, Formal analysis, Visualization, Writing -- original draft, Writing -- review \& editing. \textbf{Nicholas C. Dorn}: Software, Investigation, Data curation, Writing -- review \& editing. \textbf{Andras Lasso}: Software, Writing -- review \& editing. \textbf{Matthew A. Jolley}: Writing -- review \& editing. \textbf{Jeffrey A. Feinstein}: Resources, Supervision, Writing -- review \& editing. \textbf{Doug L. James}: Conceptualization, Supervision, Writing -- review \& editing. \textbf{Alison L. Marsden}: Supervision, Funding acquisition, Writing -- review \& editing.

%% =====================================================================
%%  DECLARATION OF COMPETING INTEREST (required by CMPB / Elsevier)
%% =====================================================================
\section*{Declaration of competing interest}

The authors declare that they have no known competing financial interests or personal relationships that could have appeared to influence the work reported in this paper.

%% =====================================================================
%%  ACKNOWLEDGEMENTS
%% =====================================================================
\section*{Acknowledgements}

Funding: This work was supported by the National Institutes of Health [grant number R01HL167516] and the National Science Foundation [grant number 2310909]. The funders had no role in the study design, data collection and analysis, decision to publish, or preparation of the manuscript. The authors would like to thank Dr Gregory T. Adamson and Dr Doff B. McElhinney for discussions, clinical knowledge, and assistance with data curation. The authors also acknowledge the Texas Advanced Computing Center (TACC) at The University of Texas at Austin for providing computational resources, supported by NSF Award \#2320757, that have contributed to the research results reported within this paper.

%% =====================================================================
%%  DECLARATION OF AI-USE
%% =====================================================================
\section*{Declaration of generative AI and AI-assisted technologies in the manuscript preparation process}

During the preparation of this work the authors used Claude (Anthropic)
to assist with language editing and improving readability. After using
this service, the authors reviewed and edited the content as needed and take full responsibility for the content of the published article.

%% =====================================================================
%%  REFERENCES
%% =====================================================================
\bibliographystyle{elsarticle-num}
\bibliography{main-reference}

\end{document}